\documentclass{jpp}
\usepackage{graphicx}

\usepackage[utf8]{inputenc}
\usepackage[T1]{fontenc}
\usepackage{amsmath}
\usepackage{color}
\usepackage[colorlinks=true, allcolors=blue]{hyperref}
\usepackage{url}

\shorttitle{Ion-acoustic turbulence}
\shortauthor{Z. Liu, R. White, M. Francisquez, L. M. Milanese and N. F. Loureiro}

\title{A Two-dimensional Numerical Study of Ion-Acoustic Turbulence}

\author{Zhuo Liu\aff{1}
  \corresp{\email{zhuol@mit.edu}},
  Ryan White \aff{2},
  Manaure Francisquez \aff{3},
  Lucio M. Milanese \aff{4}
 \and Nuno F. Loureiro \aff{1}}

\affiliation{\aff{1}Plasma Science and Fusion Center, Massachusetts Institute of Technology, Cambridge, MA 02139, USA
\aff{2}AFLCMC Directorate, Department of Defense
\aff{3}Princeton Plasma Physics Laboratory, Princeton, NJ 08543, USA
\aff{4}Proxima Fusion GmbH, Munich, Germany
}

\newcommand\bb[1]{\mbox{\boldmath{$#1$}}}

\begin{document}

\maketitle

\begin{abstract}
We investigate the linear and nonlinear evolution of the current-driven ion-acoustic instability in a collisionless plasma via two-dimensional Vlasov-Poisson numerical simulations.
We initialize the system in a stable state and gradually drive it towards instability with an imposed, weak external electric field, thus avoiding physically unrealizable super-critical initial conditions.
A comprehensive analysis of the nonlinear evolution of ion-acoustic turbulence (IAT) is presented, including the detailed characteristics of the evolution of the particles' distribution functions, (two-dimensional) wave spectrum, and the resulting anomalous resistivity.
Our findings reveal the dominance of 2D quasi-linear effects around saturation, with nonlinear effects, such as particle trapping and nonlinear frequency shifts, becoming pronounced during the later stages of the system's nonlinear evolution.
Remarkably, the Kadomtsev-Petviashvili (KP) spectrum is observed immediately after the saturation of the instability.
Another crucial and noteworthy result is that no steady saturated nonlinear state is ever reached: strong ion heating suppresses the instability, which implies that the anomalous resistivity associated with IAT is transient and short-lived, challenging earlier theoretical results.
Towards the conclusion of the simulation, electron-acoustic waves (EAWs) are triggered by the formation of a double layer and strong modifications to the particle distribution induced by IAT.
\end{abstract}

\section{\label{sec:introduction}Introduction}

A defining property of weakly collisional plasmas is their ability to support resonant energy transfer between waves and particles. The best-known manifestation of this phenomenon is the famous Landau damping, whereby waves are resonantly damped via energy transfer to the plasma particles~\citep{Landau1946}. 
A contrasting possibility is the case where energy flows in the opposite direction, from the particles to the waves, thereby growing these to nonlinear amplitudes and triggering a wide variety of complex, and often poorly understood, nonlinear plasma behavior.
Such kinetic instabilities are fundamental and ubiquitous plasma processes, thought to regulate, or at least significantly contribute to, particle behavior in weakly collisional plasmas, thereby determining their large-scale properties.
For example, it is conjectured that the whistler instability can greatly reduce the heat flux in weakly magnetized collisionless plasmas~\citep{roberg2018suppression}; and mirror and firehose instabilities are thought to regulate the ion distribution and, thus, limit the temperature anisotropy in the Earth's magnetotail~\citep{zhang2018whistler} and in solar wind~\citep{alexandrova2013solar}.

Kinetic instabilities may also be crucial in magnetic reconnection events in collisionless plasmas, because the intense wave-particle interactions that they trigger may create an anomalous resistivity \citep[e.g.,][]{Sagdeev1967, Galeev1984, Labelle1988} that breaks the frozen flux condition and potentially sets the reconnection rate \citep[e.g.,][]{ji1998experimental, kulsrud1998magnetic, kulsrud2001magnetic,treumann2001origin, uzdensky2003petschek}.
Indeed, there is observational \citep[e.g.,][]{Deng2001, Farrell2002, Matsumoto2003} and numerical ~\citep[e.g.,][]{drake2003formation, zhang2022} evidence that verifies the existence of turbulent phenomena and wave emissions around diffusion regions of reconnection sites. 
In addition, kinetic instabilities may influence the reconnection onset (i.e., the sudden transition from a relatively quiescent state to the reconnection stage proper) by disturbing the current sheet formation process and ultimately setting the properties of the reconnecting current sheet~\citep[e.g.,][]{alt2019onset, winarto2021triggering}.

As one of the most prominent examples of streaming-type kinetic instabilities, ion-acoustic instability (IAI) --- which arises when the drift velocity between electrons and (relatively cold) ions exceeds a threshold of the order of the ion sound speed --- is a strong candidate for explaining many observations in various plasma environments.
Dating from as far back as the 1970s, observations by the Helios I \& II spacecraft revealed the presence of ion-acoustic waves (IAWs) at heliocentric distances between 0.3 and 1 AU~\citep{gurnett1977plasma, gurnett1978ion}. 
Recently, IAWs have been receiving progressively more attention thanks to ongoing space missions, namely NASA's Magnetospheric Multiscale Mission (MMS)~\citep{Burch2016} and Parker Solar Probe (PSP)~\citep{fox2016solar}; and European Space Agency's Solar Orbiter~\citep{Muller2013}. 
For example, recent data from the Time Domain Sampler receiver on Solar Orbiter identifies the IAW as a dominant wave mode in the near-Sun solar wind below the local electron plasma frequency~\citep{2021pisa, graham2021kinetic}. 
Nonlinear structures in the particles' distribution functions associated with nonlinear IAWs, such as ion holes and electron holes, were also recently reported~\citep{Mozer2020a, Mozer2020b}. 
In addition, IAWs have also been identified in the separatrix and outflow region on the magnetospheric side of the reconnecting magnetopause \citep{uchino2017waves, steinvall2021large}, which further suggests that IAI may indeed be one essential component of magnetic reconnection in collisionless environments.
Understanding the nonlinear evolution of IAI, or IAT, is thus central to the interpretation of these observations. 

Attempts to understand IAT began as early as the 1950s.
Early analytical works either considered electron scattering by turbulence pulsations or induced scattering of waves by ions as the mechanisms underlying the saturation of the instability, and driving anomalous resistivity.
The former is commonly described using a quasi-linear approach that assumes a weak turbulence level and unperturbed ions~\citep{rudakov1966quasilinear,kovrizhnykh1966interaction, zavoiski1967turbulant}; whereas the latter is mainly studied semi-quantitatively based on the Kadomtsev-Petviashvili (KP) model considering wave-ion nonlinear interactions~\citep{kadomtsev1962turbulence,petviashvili1963ion}.
In 1969, R.~Z. Sagdeev derived an expression for IAT-induced anomalous resistivity by calculating the nonlinear wavenumber spectrum resulting (exclusively) from the nonlinear scattering of waves by ions~\citep{Sagdeev1967, sagdeev1969nonlinear}. 
Although Sagdeev's resistivity formula has been widely adopted~\citep[e.g.,][]{uzdensky2003petschek}, it is important to note that it is a partially qualitative result, because it fails to consider the angular distribution of IAT.
Additionally, its general validity is unclear because the scattering of waves by ions is not the only mechanism determining the nonlinear evolution of IAI:
changes in the particles' velocity distribution as the waves grow and saturate can in principle be just as important and are not considered in Sagdeev's derivation.
A comprehensive quantitative nonlinear theory of IAT was not developed until the 1980s when V. Yu. Bychenkov and colleagues solved the kinetic equations analytically. 
Their work simultaneously accounted for the quasi-linear interaction of electrons with waves and the induced scattering of waves by ions, enabling the quantitative analysis of anomalous turbulent transport~\citep{bychenkov1981angular, bychenkov1982ion, bychenkov1982transfer, bychenkov1984transport}. 
By establishing the spectral and angular distribution of ion-acoustic turbulent pulsations, they verified the correctness of Sagdeev's resistivity formula in cases where the scattering of waves by ions dominates the saturation mechanisms.

Despite these achievements, significant uncertainties remain.
First, many assumptions or approximations are made in the above-mentioned theories, often excluding possibilities that may be crucial in determining the dynamics of the system. 
For example, nearly all analytical models assume that a nonlinear steady state will be reached, which is not necessarily true in a realistic situation (and will indeed not be the case in this study).
Second, no theory properly considers how the modification of the ion velocity distribution due to the heating of resonant ions changes the IAT. Thus, numerical simulations are needed to identify the key nonlinear mechanisms and either validate the current theoretical understanding or, instead, guide the development of a new, improved one.

Numerical studies of IAT started in the 1970s, with most simulations performed using the particle-in-cell (PIC) method~\citep[e.g.,][]{biskamp1971computer,biskamp1972computer,boris1970computations,ishihara1981quasilinear,ishihara1983quasilinear,dum1979anomalous,degroot1977localized}.
Most of these studies focused on identifying the saturation mechanism of IAT and quantifying the resulting turbulent heating.
However, rather than gradually driving the system toward an unstable state --- which is the scenario that better conforms to the theoretical approaches to this problem --- these simulations start with super-critical initial conditions (their initial condition is such that the electron-ion drift velocity significantly exceeds the threshold for IAI); in some cases, 
the electron drift velocity is even kept constant throughout the simulation.
As a result, they are inadequate to directly verify the analytical theories.
Moreover, the relatively small numbers of macro-particles employed in those PIC simulations (due to computational constraints) produced unresolved results. 
Indeed, even with today's computational resources, simulating resonant kinetic instabilities with the PIC method can be challenging (see, e.g., ~\citet{tavassoli2021role} for an example concerning the Buneman instability). 

The availability of efficient continuum Vlasov-Poisson solvers which arose at the turn of the century~\citep[e.g.,][]{fijalkow1999numerical,horne2001new} enabled a resurgence of numerical studies of IAT.
These new continuum Vlasov-based simulations systematically studied the relationship between anomalous resistivity and different simulation parameters such as different types of initial distribution functions~\citep{watt2002ion, petkaki2003anomalous}, the initial drift velocity of electrons~\citep{petkaki2008nonlinear}, and phase space structures ~\citep{buchner2006anomalous,lesur2014nonlinear}. 
Some of these simulations adopted realistic electron-ion temperature and mass ratios~\citep[e.g.,][]{petkaki2006anomalous,hellinger2004effective}. 
One conclusion from most of these numerical studies was that the anomalous resistivity created by IAT was at least an order of magnitude larger and more strongly dependent on the drift velocity of electrons than the theoretical estimation~\citep{galeev1984current}. 

However, despite the valuable insights provided by these simulations, some potential problems persist. 
One is related to the starting configurations used in the simulations, which are often super-critical. 
It remains unclear whether such initial configurations are achievable in realistic scenarios, as their realization depends on the relative rates of current development and instability growth.
Another concern is that, due to computational constraints, these simulations are restricted to one-dimension in both real and velocity space (1D1V), whereas the importance of two-dimensional (2D) effects has been identified in the early numerical work~\citep{biskamp1971computer}.
Analytical studies~\citep[e.g.][]{zavoiski1967turbulant,sagdeev1969nonlinear} also considered the influence of oblique modes on the instability dynamics.
Perhaps as a result of this artificial restriction, the nonlinear evolution observed in these simulations is mostly highly stochastic. 
Consequently, none of these studies show a direct comparison against the theoretical spectra of IAT pulsations and Sagdeev's formula for anomalous resistivity. 

These limitations mean that considerable uncertainty remains about the nonlinear properties of the IAI; in particular, fundamental questions such as its saturation mechanism, the resulting particle heating, and the magnitude of the anomalous resistivity that it drives remain unanswered. 
In this paper, we conduct a comprehensive numerical study of the nonlinear evolution of the current-driven IAI aimed at providing detailed answers to these questions. 
We numerically solve the Vlasov-Poisson equations in two dimensions both in real and in velocity space.
To ensure the physical realizability of our system, our simulations start from a stable initial condition which is driven gradually towards instability via an imposed external electric field. 
This setup allows us to present the most complete and detailed understanding of IAT to date.

This paper is organized as follows.
Numerical details are provided in section~\ref{sec:setup}. 
The results from our main simulation are presented and discussed in section~\ref{sec:results_1}. 
In section~\ref{sec:results_2}, we analyze the sensitivity of our results to the amplitude of the external electric field, the electron-ion mass ratio,  and the initial temperature ratio. Lastly, conclusions from our work and a discussion of the implications of our results in broader contexts are presented in section~\ref{sec:conclusion}.

\section{\label{sec:setup}Numerical Setup}
We investigate IAI in a uniform, spatially 2D plasma with periodic boundary conditions (thus effectively mimicking an infinitely large 2D plasma).
All simulations start with uniform Maxwellian electrons and ions, with zero drift velocity (i.e., no net current).
We focus on the classic IAI and, therefore, consider only the case when electrons are initially much hotter than the ions ($T_{e0}/T_{i0} \gg 1 $).
An external electric field $E_{\rm{ext}}$ is applied in the $\bb{z}$ (called parallel) direction throughout the simulation to drive the current and, thus, the instability. 
The reference value against which $E_{\rm{ext}}$ must be compared is~\citep{bychenkov1988ion}
\begin{equation}
    \label{eq:E_NL}
    E_{\rm{NL}} = \frac{m_e c_{s0} \omega_{pi} T_{e0}}{6\pi e T_{i0}},
\end{equation}
where $c_{s0} = \sqrt{T_{e0}/m_i}$ is the ion sound speed, $m_e$ is the electron mass, $\omega_{pi}$ is the ion plasma frequency, and $\lambda_{De}$ and $\lambda_{Di}$ are the electron and ion Debye lengths. 
In this work, we focus on the so-called weak field case, $E_{\rm{ext}} \ll E_{\rm{NL}}$. 
This choice is guided by simplicity considerations: in this regime, wave-wave interactions are predicted to be sub-dominant, and quasi-linear theory is expected to apply~\citep{bychenkov1988ion}. 
This is, therefore, a necessary first step in the development of a complete understanding of IAT. 

We use {\tt Gkeyll} ~\citep{shi2017gyrokinetic, shi2019full, mandell2020electromagnetic,Juno:2018,HakimJuno:2020}, a recently developed state-of-the-art code featuring energy-conserving, high-order discretization methods to solve the 2D Vlasov-Poisson equations:
\begin{equation}
 \begin{aligned}
\frac{\partial f_{\alpha}}{\partial t}+\mathbf{v} \cdot \nabla f_{\alpha}+ \left(\mathbf{E_{\rm{ext}}} -  \frac{q_{\alpha}}{m_{\alpha}} \nabla \varphi \right) \cdot \frac{\partial f_{\alpha}}{\partial \mathbf{ v}}=0, \\
\nabla^{2} \varphi=-4 \pi \sum_{\alpha} q_{\alpha} \int \mathbf{d}^{3} v f_{\alpha}.
 \end{aligned}
 \label{eq:vlasov-possion}
\end{equation}. 

\begin{table}
\begin{center}
\def~{\hphantom{0}}
\renewcommand\arraystretch{1.2}
\setlength{\tabcolsep}{12pt}
\caption{\label{tab:sim_summary} Summary of the key parameters of the simulations. In the run name, the number following the letter {\tt M} refers to the mass ratio. The number following the letter {\tt E} refers to the multiple of $2.5\times10^{-4}$ for $\tilde{E}_{\rm ext}$.}
\begin{tabular}{ccccc}
\hline
Run & $\tilde{E}_{\rm{ext}}$ & $E_{\rm NL} / (4\pi en_0\lambda_{De})$ & $m_i/m_e$  &  $T_{e0}/T_{i0}$ \\ \hline
{\tt Main}   & $2.5\times 10^{-3}$ & $2.7\times 10^{-2}$  & $100$ & $50$    \\
{\tt M25E10}  & $2.5\times 10^{-3}$ & $1.1\times 10^{-1} $  & $25$  & $50$    \\
{\tt M50E10}  & $2.5\times 10^{-3}$ & $5.3 \times 10^{-2}$  & $50$  & $50$    \\
{\tt M200E10} & $2.5\times 10^{-3}$ & $1.3 \times 10^{-2}$  & $200$ & $50$    \\
{\tt M25E1}  & $2.5\times 10^{-4}$ & $1.1 \times 10^{-1}$  & $25$  & $50$    \\
{\tt M25E2}  & $5.0\times 10^{-4}$ & $1.1 \times 10^{-1}$  & $25$  & $50$    \\
{\tt M25E4}  & $1.0\times 10^{-3}$ & $1.1 \times 10^{-1}$  & $25$  & $50$    \\
{\tt M25E6}  & $1.5\times 10^{-3}$ & $1.1 \times 10^{-1}$  & $25$  & $50$    \\
{\tt M25E8}  & $2.0\times 10^{-3}$ & $1.1 \times 10^{-1}$  & $25$  & $50$    \\
{\tt T20}    & $2.5\times 10^{-3}$ & $4.2 \times 10^{-2}$  & $25$  & $20$    \\
{\tt T100}   & $2.5\times 10^{-3}$ & $2.1 \times 10^{-1}$  & $25$  & $100$   \\
\hline
\end{tabular}
\end{center}
\end{table}

We perform a set of simulations varying the magnitude of the external electric field (restricted to the weak electric field regime), the mass ratio, and the initial temperature ratio. 
For convenience, we introduce the normalized external electric field $\tilde{E}_{\rm{ext}} \equiv  E_{\rm ext}/ (4\pi e n_0 \lambda_{De})$, where $\lambda_{De} = v_{Te0}/\omega_{pe}$ is the electron Debye length corresponding to the initial electron temperature.
Table~\ref{tab:sim_summary} lists all simulations and the corresponding values of $\tilde{E}_{\rm{ext}}$, $E_{\rm NL}$, mass ratio, and $T_{e0}/T_{i0}$.
Run {\tt Main} is our fiducial simulation, with a mass ratio $m_i/m_e=100$, temperature ratio $T_{e0}/T_{i0}=50$, and an external electric field of $\tilde{E}_{\rm{ext}} =  2.5\times 10^{-3}$. 
Most other runs vary the mass ratio or electric field at fixed $T_{e0}/T_{i0}=50$. 
In the run name, the number that follows the letter {\tt M} refers to the mass ratio, whereas the number following {\tt E} refers to the multiple of $2.5\times 10^{-4}$ for $\tilde{E}_{\rm{ext}}$.
The last two runs listed in table~\ref{tab:sim_summary} are meant to investigate the effect of the initial temperature ratio.

For all simulations, the spatial domain size is $50.0 \lambda_{De}$ in the $\bb{z}$ direction and $25.0 \lambda_{De}$ in the $\bb{y}$ direction, with grid resolution $\Delta z = \Delta y = 0.5\lambda_{De}$.
The resolution of electron and ion velocity space grids are jointly determined by linear theory, linear benchmarking, and a series of nonlinear evolution tests and, therefore, vary among the different simulations. 
We note, however, that simulation  {\tt M200E10} is performed at numerical resolutions that cannot resolve the linear stage of ion-acoustic modes with small wavelengths (but it can resolve the most unstable linear mode), due to limited computational resources. 
However, according to our nonlinear evolution tests, this shortcoming has only a very limited impact on the evolution of the current and heating in the nonlinear stage.
A detailed description of how we determine spatial and velocity space resolution is provided in appendix~\ref{app:5}.

In section~\ref{sec:results_1}, we use our fiducial simulation, Run {\tt Main}, to demonstrate our main results.
The other simulations are employed in section~\ref{sec:results_2} to discuss how the results depend on the parameters listed in table~\ref{tab:sim_summary} and to extrapolate our results to the realistic mass-ratio case.


\section{\label{sec:results_1} IAT driven by a weak electric field}

In this section, we report on Run {\tt Main}, for which we plot in Figure~\ref{fig:1.current_and_energy} the time traces of electron current in the $\bb{z}$ direction and total wave energy.
Based on this figure, we identify five distinct phases of evolution, which we analyze in detail in the subsections below.

\subsection{Phase I: linear stage}
At the start of the simulation, both electrons and ions are at rest.
They are freely accelerated by the external electric field, thus ramping up the current and driving the system toward instability. 
During this stage, the current growth rate is the free acceleration rate, i.e.,  $dJ_z/dt = E_{\rm{ext}} e^2 n_0 / m_e$ (which neglects the small ion contribution to the current), where $J_z$ is the parallel current in the system.

\begin{figure}
\centering
\includegraphics[width=0.7\textwidth]{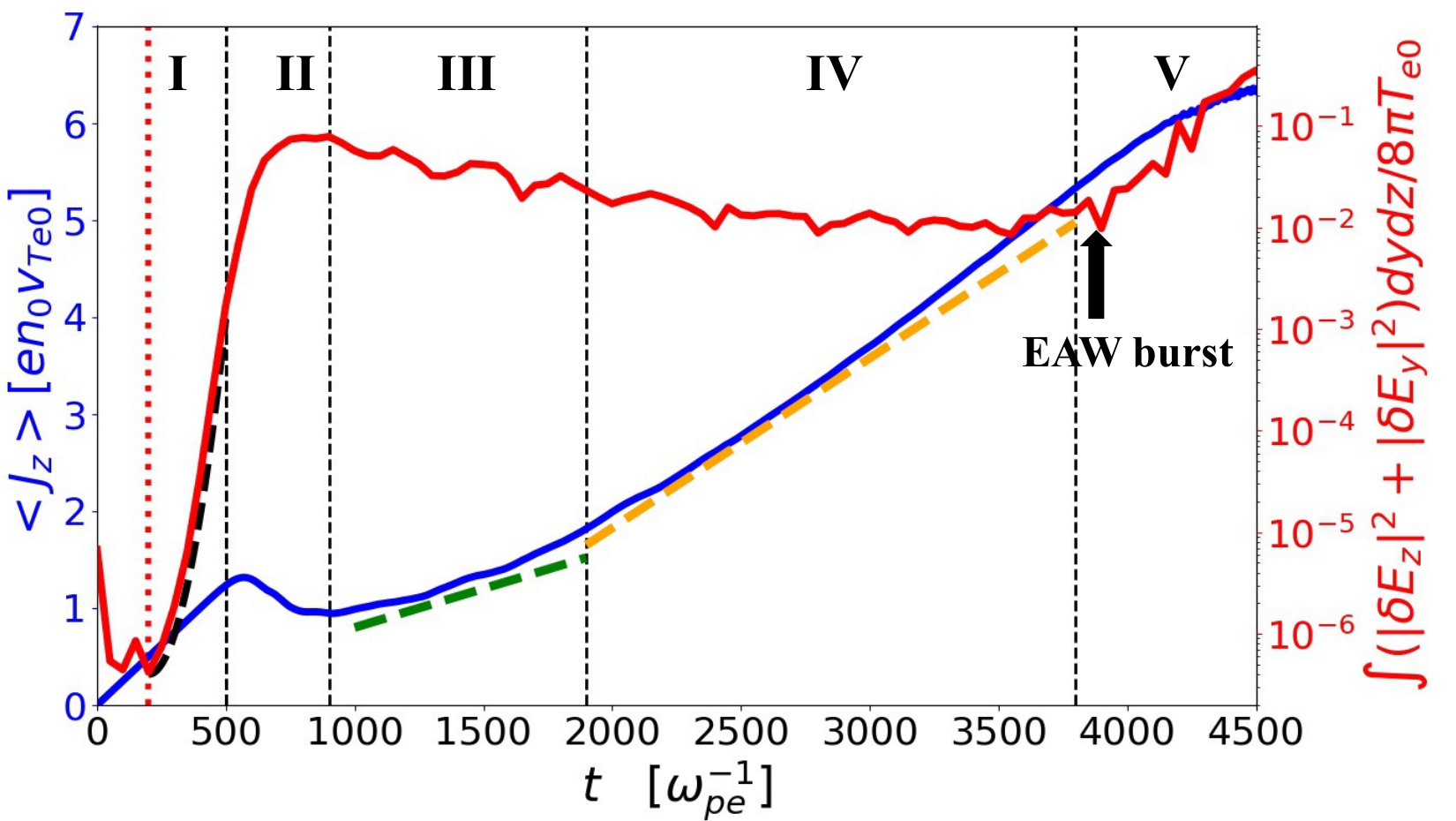}
\caption{\label{fig:1.current_and_energy} Time trace of total wave energy (red curve, right axis) and parallel current (blue curve, left axis) of Run {\tt Main}.  We split the evolution of the system into five different stages; see text for details. The red dotted vertical line indicates the moment when the current growth rate is comparable to that of the most unstable IAW predicted by linear theory. The black dashed line is $W(t)$ predicted by linear theory, see Eq.~\eqref{eq:linear_fit}. The green dashed line indicates the current growth rate during phase III (which is 30\% of the free acceleration rate), and the orange dashed line indicates the current growth rate during phase IV (which is 85\% of the free acceleration rate).}
\end{figure}

In Figure~\ref{fig:1.current_and_energy}, the red dotted vertical line indicates the moment of time, $t_0$, when the current growth rate ($1/J_z\, dJ_z/dt$) is comparable to that of the most unstable IAW predicted by linear theory~\citep{jackson1960drift} (linear theory is briefly described in appendix~\ref{app:0}). 
We observe that this criterion predicts the onset of the instability very well.
According to linear theory, the growth rate of an ion-acoustic mode, $\gamma_e(k)$, is roughly proportional to the drift velocity of electrons (as long as the drift velocity of the electrons is much larger than the ion sound speed). Because the electrons are freely accelerated by the external electric field, the linear growth rate of an ion-acoustic mode at time $t$ is then approximately $\gamma_e(t, k) \approx \gamma_e(t_0, k)(t/t_0)$.
Therefore, the wave amplitude should grow as
\begin{equation}
\label{eq:linear_fit}
    W(t) = W(t_0) e^{2\gamma_e (t,k) t} \approx W(t_0)  e^{\frac{2\gamma_e(t_0,k)}{t_0} t^2}.   
\end{equation}
We calculate $\gamma_e(t_0, k)$ of the most unstable mode using linear theory, and the resulting $W(t)$ is plotted with the black dashed line in Figure~\ref{fig:1.current_and_energy}. The observed wave energy agrees reasonably well with Eq.~\eqref{eq:linear_fit}.

This linear phase ends when the wave energy becomes large enough that wave-particle interaction becomes significant and the electron velocity distribution begins to be modified, at around $\omega_{pe}t \approx 500$.

\subsection{\label{sec:phase2}Phase II: saturation}

\begin{figure}
\centering
\includegraphics[width=1.0\textwidth]{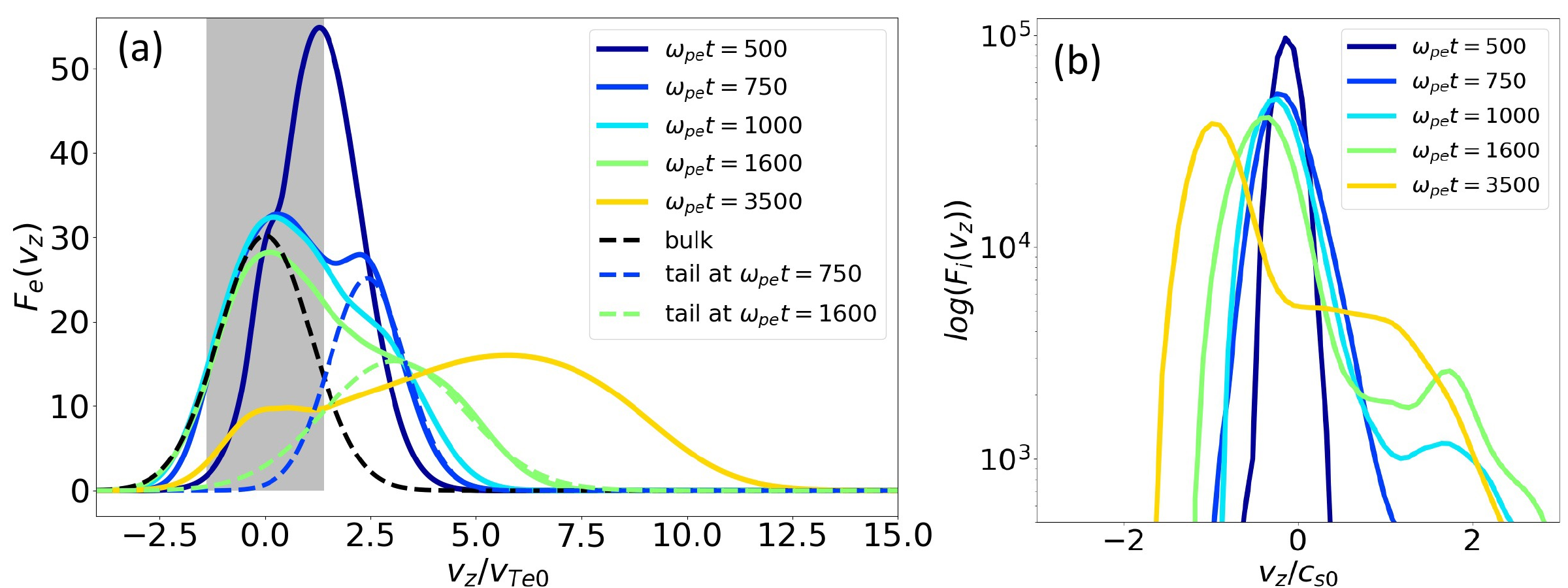}
\caption{\label{fig:2.dist_1d} 1D electron (a) and ion (b) (in logarithmic scale) velocity distribution at different times. The shadowed region is the approximate electron resonance region at the moments around saturation ($\omega_{pe} t \approx 750$). The black dashed line represents the fitted bulk in the electron velocity distribution after saturation, and the dashed blue and green curves represent the fitted tail at different times. The fitting model is the double Maxwellian function, Eq.~\eqref{eq:double_max}. See text for more details.} 
\end{figure}

\begin{figure}
\centering
\includegraphics[width=0.6\textwidth]{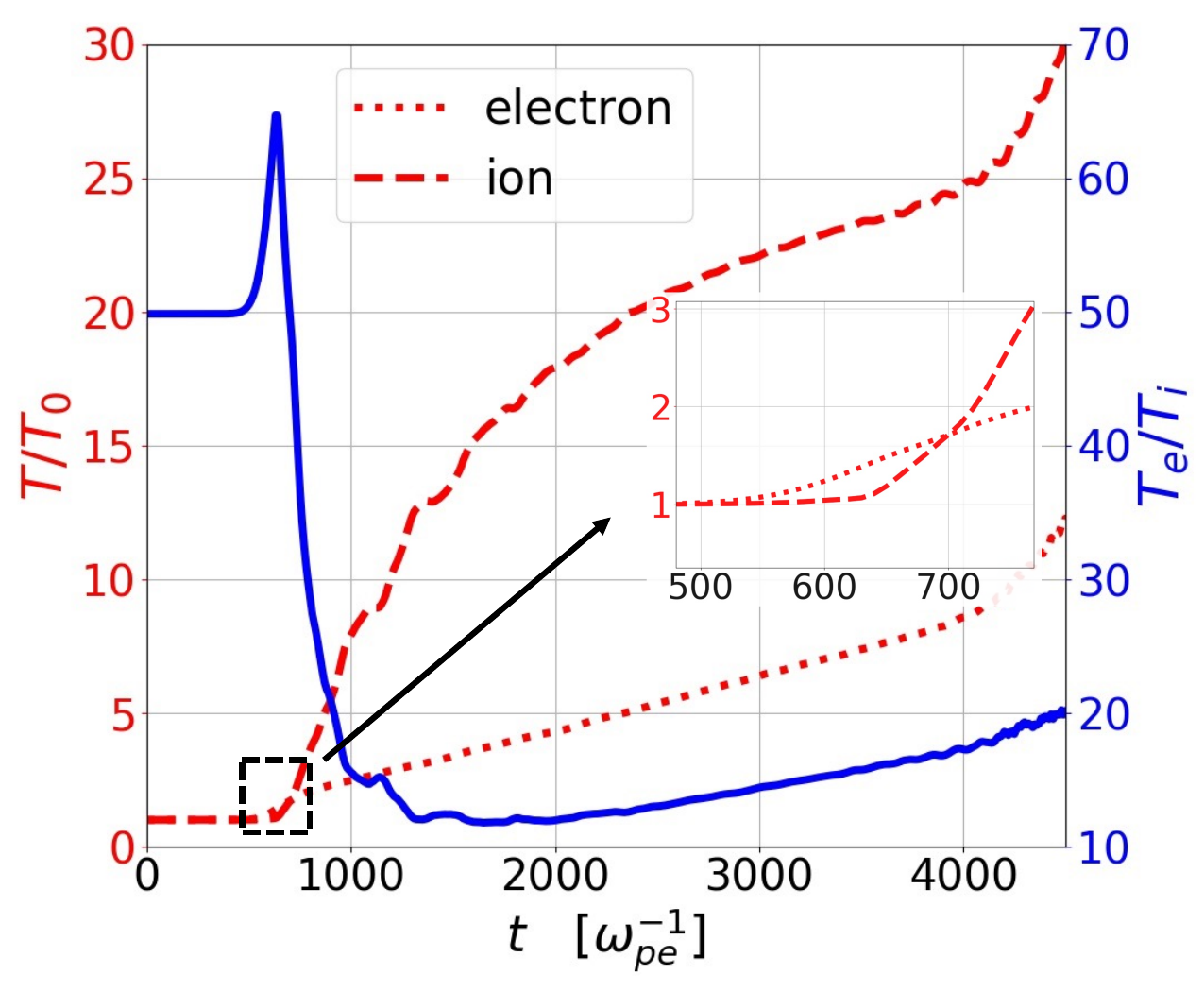}
\caption{\label{fig:3.temp} Time traces of electron and ion temperatures, normalized to their respective initial temperature (red dashed and red dotted lines, left axis), and time trace of electron-to-ion temperature ratio (blue line, right axis). The temperatures are defined as the second moment of their own velocity distribution functions.}
\end{figure}

Phase II captures the saturation stage of IAI. 
We first investigate the saturation mechanism. 
As mentioned in section~\ref{sec:introduction}, two processes are potentially responsible for saturation: quasi-linear interactions between electrons and waves, and nonlinear interactions with ions (ion trapping ~\citep{sagdeev1969nonlinear,biskamp1971computer}). 
The electrons' and ions' responses to the growth of the IAWs are represented in Figures~\ref{fig:2.dist_1d} (a) and (b), respectively, by plotting the one-dimensional (1D) velocity distribution function at different moments of time, obtained by averaging over the spatial domain ($z,y$) and $v_y$; namely, $F_{\alpha}(v_z) \propto \int f_{\alpha}(v_z,v_y, z,y) dz\, dy\, d v_y $. 
As the dark blue curve in Figure~\ref{fig:2.dist_1d}(a) shows, a quasi-linear plateau starts forming in the resonance region in the electron velocity distribution at around $\omega_{pe}t\approx 500$, which weakens the growth of the IAI.
This moment of time coincides with the onset of strong electron heating, as evidenced by the electron temperature time trace shown in the inset plot of Figure~\ref{fig:3.temp}.
Importantly, we observe (same figure) that strong ion heating (which is the signature of ion trapping) only begins later in time, at around $t\approx 650\, \omega_{pe}^{-1}$, when the growth rate of wave energy is already significantly reduced and the wave energy is about to arrive at its peak. 
This suggests that ion trapping is \textit{not} the main saturation mechanism.
We further verify this conjecture by summarizing the maximum wave energy density we observe in simulations with different mass ratios and external electric fields $W_{\rm{max}}$ in table~\ref{tab:saturation} and comparing them with the quasi-linear estimate~\citep{zavoiski1967turbulant, bychenkov1988ion}, 
\begin{equation}
\label{eq:W_sat}
    W_{\rm{sat}} \approx n_0T_{e0} \tilde{E}_{\rm{ext}}\left(\frac{m_i}{2m_e}\right)^{1/2},
\end{equation}
where $W_{\rm{sat}}$ is the theoretical wave energy density at saturation.
Our numerical results indeed show an approximately linear dependence of the peak wave energy on the external electric field $E_{\rm{ext}}$ and on the square root of mass ratio $(m_i/m_e)^{1/2}$.
In addition, the absolute value of maximum wave energy density is only a little bit larger than Eq.~\eqref{eq:W_sat}.
Therefore, we conclude that, under the weak external electric field condition, quasi-linear relaxation of the electron distribution, rather than ion trapping, is the main IAI saturation mechanism, consistent with the theoretical prediction~\citep{bychenkov1988ion}. 
This conclusion is seemingly at odds with the numerical study by \citet{biskamp1971computer}, which argues instead that ion trapping is the mechanism responsible for saturation.
We think that this discrepancy is due to the fact that their simulations start with a super-critical electron distribution. 
This causes the waves to grow rapidly to large amplitudes. 
Thus, nonlinear effects, rather than quasi-linear relaxation of electron distribution, dominate the saturation of the IAI in their simulations.

\begin{table}
\begin{center}
\renewcommand\arraystretch{1.2}
\setlength{\tabcolsep}{12pt}
\begin{tabular}{lcc}
\hline
Run & $W_{\rm{sat}}/nT_{e0}$ (\%) & $W_{\rm{{max}}}/nT_{e0}$ (\%) \\
\hline
 {\tt M25E2} & 0.18 & 0.19\\
 {\tt M25E4} & 0.35 & 0.40\\
 {\tt M25E10} & 0.88 & 1.06\\
 {\tt Main}  & 1.77 & 1.91\\
 {\tt M200E10} & 2.50 & 2.62\\
 \hline
\end{tabular}
\caption{ The saturation wave energy of simulations with different mass ratios and external electric fields.  $W_{\rm{max}}$ is the maximum wave energy measured from the simulation. $W_{\rm{sat}}$ is the quasilinear estimate of the saturated wave energy, i.e.,  Eq.~\eqref{eq:W_sat}.
The wave energy exhibits approximate linear dependence on the external electric field and on the square root of mass ratio, consistent with the quasi-linear prediction. }
\label{tab:saturation}
\end{center}
\end{table}

Despite a good agreement between theory and simulations on wave energy at the moment of saturation, the quasi-linear theory assumes that IAT will reach a steady state (i.e., approximately constant wave energy) determined by the balance between wave emission ($\gamma_e(k)$; weakened due to plateau formation) and wave damping ($\gamma_i(k)$), namely, 
\begin{equation}
\label{eq:quasi_linear}
    \gamma(\bb{k}) = \gamma_e(\bb{k}) + \gamma_i(\bb{k}) = 0.
\end{equation}
However, it can be seen from Figure~\ref{fig:1.current_and_energy} that a steady state is not achieved: the wave energy quickly starts to drop after saturation.
We show later in this section that Eq.~\eqref{eq:quasi_linear} is indeed achieved at the moment of saturation, but does not remain valid afterward.

To understand the nonlinear evolution after saturation, observe the blue ($\omega_{pe}t=750$) and the cyan ($\omega_{pe}t=1000$) curves in Figure~\ref{fig:2.dist_1d}(a). 
They display two noteworthy features: (i) the (1D) electron velocity distribution does not retain the quasi-linear plateau at the end of the saturation process; (ii)
the non-resonant part is mostly represented by a tail (see the blue dashed curve in Figure~\ref{fig:2.dist_1d}(a)), which appears due to significant electron heating (see Figure~\ref{fig:3.temp}) and acceleration by the external electric field during saturation.
The disappearance of the plateau shape in the electron velocity distribution is, in fact, attributed to the saturation of oblique modes, which is a notable 2D feature. 
The much more efficient scattering of electrons by the waves in the 2D situation brings the majority of electrons back to the resonance regime. More evidence of this can be found in appendix~\ref{app:1}.

Based on these observations, we find that the 1D electron velocity distribution at the end of, and after, saturation can be well approximated by a double Maxwellian function, with one Maxwellian representing the bulk of the electron population (denoted with subscript `b' in the following formula), and the other representing the tail (subscript `t'):
\begin{align}
\label{eq:double_max}
\begin{split}
    F_e(v_z) \approx (1-\alpha) & \frac{1}{\sqrt{2\pi}v_{Te,\rm{b}}} \exp \left(-\frac{(v_z-u_{d, \rm{b}})^2}{2v_{Te,\rm{b}}^2}\right) +\\
\alpha & \frac{1}{\sqrt{2\pi}v_{Te,\rm{t}}} \exp \left(-\frac{(v_z-u_{d,\rm{t}})^2}{2v_{Te,\rm{t}}^2}\right).
\end{split}
\end{align}
This bulk-tail model is predicated on the notion that, as the wave energy saturates, a resonant and a non-resonant part should be manifest in the electron velocity distribution.
The acceleration of the resonant part by the external electric field should be countered by the anomalous resistivity created by IAT, while the non-resonant part can still be freely accelerated. 
The fitted distribution at the moment of saturation ($\omega_{pe}t \approx 750$) is plotted with a black dashed curve (the bulk) and a dashed blue curve (the tail) in Figure~\ref{fig:2.dist_1d}(a). The exact fitting parameters can be found in appendix~\ref{app:2}.

Using the empirical fit, we now demonstrate that Eq.~\eqref{eq:quasi_linear} still holds at the end of the saturation even though the electron distribution no longer conforms to the plateau shape described by the standard quasi-linear theory (a similar method was previously employed in studies of Buneman instability in reconnection layers~\citep{drake2003formation, che2009nonlinear, che2010electron, jain2011modeling}). 
Because ions do not deviate significantly from a Maxwellian distribution around saturation (see Figure~\ref{fig:2.dist_1d}(b)), the 1D dielectric function obtained from linear theory can be written as
\begin{equation}
\label{eq:dielectric}
    1+ \frac{1+\zeta_{i} Z\left(\zeta_{i}\right)}{k_z^{2} \lambda_{D i}^{2}}  
    +(1-\alpha)\frac{1+\zeta_{e,\rm{b}} Z\left(\zeta_{e,\rm{b}}\right)}{k_z^{2} \lambda_{D e,\rm{b}}^{2}} 
    + \alpha \frac{1+\zeta_{e,\rm{t}} Z\left(\zeta_{e,\rm{t}}\right)}{k_z^{2} \lambda_{D e,\rm{t}}^{2}} = 0,
\end{equation}
where $\lambda_{D\alpha} = v_{T\alpha}/\omega_{p\alpha}$ is the Debye length of species $\alpha$ (ions, bulk electrons or tail electrons), $\zeta_{\alpha} \equiv (\omega - ku_{d\alpha}) / kv_{T\alpha}$, with $\omega = \omega_{r} + i \gamma_e$, $u_{d\alpha}$ the drift velocity, and $Z(\zeta)$ the plasma dispersion function.
Using the fitting model provided by Eq.~(\ref{eq:double_max}) evaluated at $\omega_{pe} t = 750$,  we obtain from this equation $\gamma(k_z,\omega_{pe}t = 750) \approx 0$ for nearly all values of $k_z$ that are originally unstable.
While not strictly rigorous --- because we only consider waves in the parallel direction in Eq.~\eqref{eq:dielectric} --- this result strongly suggests that the saturation process of IAI indeed tends to bring the system to a marginally stable state as described by Eq.~\eqref{eq:quasi_linear}. 
We note that, as mentioned earlier, Eq.~\eqref{eq:quasi_linear} is not valid at later times, which voids the assumption made by the quasi-linear theory; see section~\ref{sec:phase3}.

To summarize, during the saturation process in the 2D case, electrons are efficiently scattered back to lower velocities by both parallel and oblique wave modes within the resonance region.
Consequently, the electrons within the resonance region become the dominant population, forming a new electron bulk.
After saturation, a balance is established between the effective friction provided by the waves and the acceleration from the external electric field, leading to the stationary behavior of the new bulk, while electrons located on the right-hand side of the resonance region experience continuous acceleration.

\subsection{\label{sec:phase3}Phase III: shutdown of IAT}

Phase III is a post-saturation stage characterized by a significantly lower current growth rate (about a factor of 3 lower than the free acceleration rate); see Figure~\ref{fig:1.current_and_energy}.
As explained in section~\ref{sec:phase2}, the growth of the parallel current after saturation is primarily contributed by this fitted tail.
We can estimate the current growth rate in the early stages of phase III to be approximately $\alpha(t = 1000) e^2E_{\text{ext}}/m_e$, where $\alpha(t = 1000)$ represents the fraction of the fitted tail at $t = 1000$. 
This approximation yields a value close to $0.38 e^2E_{\text{ext}}/m_e$, which is only slightly higher than the current growth rate indicated by the green dashed line in Figure~\ref{fig:1.current_and_energy}.
During phase III, the wave energy gradually decreases and reaches a level approximately one order of magnitude smaller than its peak value. 
Eventually, the intensity of the IAT becomes insufficient to provide enough friction to the bulk electrons.

\subsubsection{\label{sec:phase3-1}Landau damping induced by strong ion heating}

It has long been conjectured that the IAI will be switched off due to the reduction in the electron-to-ion temperature ratio.
This reasoning is based on the observation, from linear theory, that when $J_z/(e n)\gg c_s$ the ion-acoustic mode becomes stable for $T_e/T_i\lesssim 10$~\citep{papadopoulos1977review,benz2012plasma}.
We confirm this prediction by plotting the temperature ratio, in Figure~\ref{fig:3.temp}. 
Even though both electrons and ions are strongly heated after saturation, we observe an abrupt decrease in the electron-to-ion temperature ratio, which happens mostly in phase II and early phase III.
This plummeting of the temperature ratio can be directly responsible for a strong damping rate of IAWs for most wave modes.
\begin{figure}
\centering
\includegraphics[width=1.0\textwidth]{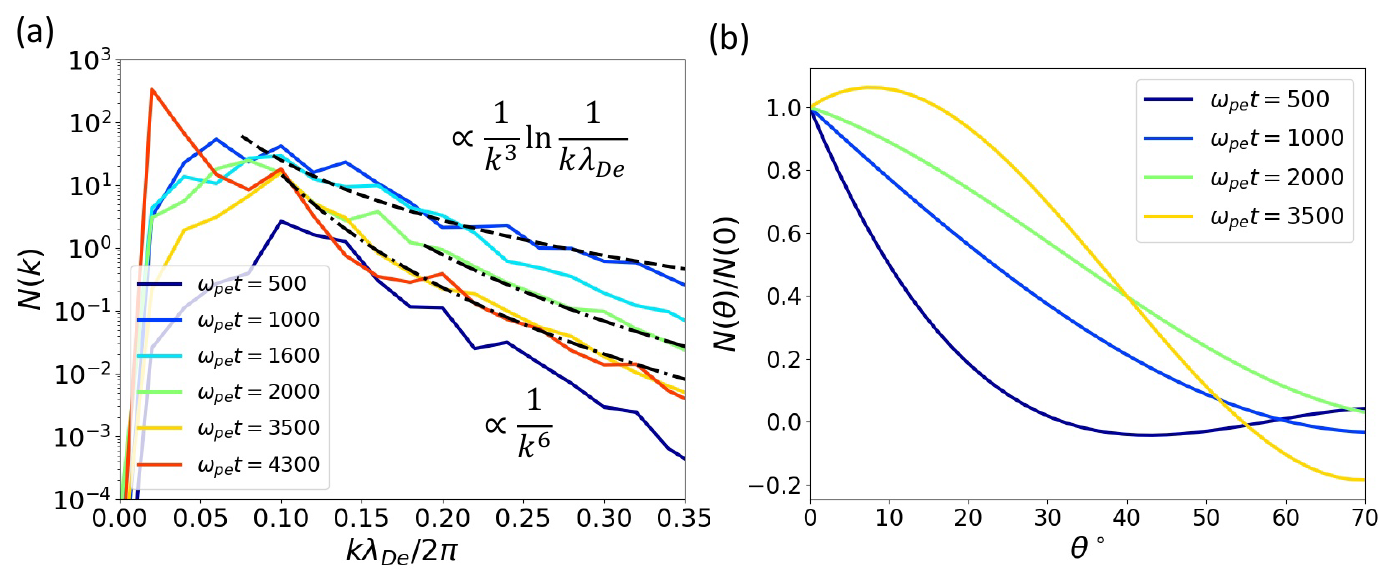}
\caption{\label{fig:4.spectrum} Wavenumber (a) and angular (b) spectra of IAWs at different times during the simulation. Angular spectra at different times are normalized to their value at $\theta = 0^{\circ}$, where $\theta$ is the angle between the parallel direction and wave propagating direction. A fourth-order polynomial is used to fit the angular spectra to smooth the curves. The wavenumber spectrum at $\omega_{pe}t=1000$ agrees with the Kadomtsev-Petviashvili (KP) spectrum (the dashed curve). The wavenumber spectra at $\omega_{pe}t=2000$ and $\omega_{pe}t=3500$ with relatively larger wavenumbers agree with the spectrum observed in~\citet{chapman2014new} (the dash-dotted curves).}
\end{figure}

However, linear theory assumes Maxwellian electron and ion velocity distributions, which is not the case in the nonlinear evolution of IAI at this stage.
To prove this conjecture more rigorously, we return to the linear analysis of the modified distribution function to show that IAWs will indeed be Landau damped after saturation.
By plugging the fitting parameters at $\omega_{pe}t=1000$ into Eq.~(\ref{eq:dielectric}) and assuming Maxwellian ions, we observe $\gamma(k_z \geq k_c,\omega_{pe}t=1000) <  - 3 \times 10^{-3} \omega_{pe}$, where $k_c \lambda_{De}/2\pi \approx 0.15$ is the smallest wavenumber to satisfy this condition ($3 \times 10^{-3} \omega_{pe}$ is the inverse of about one-third time duration of phase III).
This result implies that we should see about an order of magnitude decrease in wave energy in $\sim 300\omega_{pe}^{-1}$ for modes with wavenumber satisfying $k_z \geq k_c$ during phase III.
By comparing the wavenumber ($k$) spectrum between $\omega_{pe}t=1000$ (blue curve) and $\omega_{pe}t=1600$ (cyan curve) plotted in Figure~\ref{fig:4.spectrum}(a), we indeed see a significant decrease in amplitude for the wave modes with larger wavenumbers ($k\lambda_{De}/2\pi \gtrsim 0.2$). 
However, these wave modes are only about an order of magnitude weaker than those in $\omega_{pe}t = 1000$, which are less damped than predicted.
Using the fitting parameters at $\omega_{pe}t=1600$ in Eq.~\eqref{eq:dielectric}, we obtain $\gamma(k_z,\omega_{pe}t=1600) < 0$ for nearly all values of $k_z$, while the damping rates of wave modes with $k_z \lambda_{De}/2\pi > 0.08$ are strong (again, compared to $- 3\times 10^{-3}\omega_{pe}$).\
Even though most wave modes are less damped than predicted, the fact that nearly all the wave modes at $\omega_{pe}t=2000$ (the green curve in Figure~\ref{fig:4.spectrum}(a)) show significantly lower energy confirms our analysis.
The exact fitting parameters of the bulk-tail double Maxwellian model and more details on this linear analysis can be found in appendix~\ref{app:2}.

\subsubsection{Effects of particle trapping}
An important mechanism affecting the wave energy spectrum not considered by linear or weak-turbulence theory is particle trapping, which can strongly weaken Landau damping.
The formation of electron phase-space vortexes (also referred to as ``phase-space holes'' or ``electrostatic solitary waves'' ~\citep{hutchinson2017electron}), which is the signature of electron trapping, can be identified right after saturation (a snapshot of electron phase space at $\omega_{pe}t = 2000$ is shown in Figure~\ref{fig:6.2d}).
This phenomenon is expected by nonlinear theories, and it is also reported in many simulations of two-stream instability \citep[e.g.,][]{morse1969one, berk1970phase}, space observations where electrostatic streaming waves are present~\citep[e.g.,][]{malaspina2013electrostatic, pickett2008furthering, Mozer2020b}, and laboratory experiments~\citep[e.g.][]{saeki1979formation}. 

\begin{figure}
\includegraphics[width=1.0\textwidth]{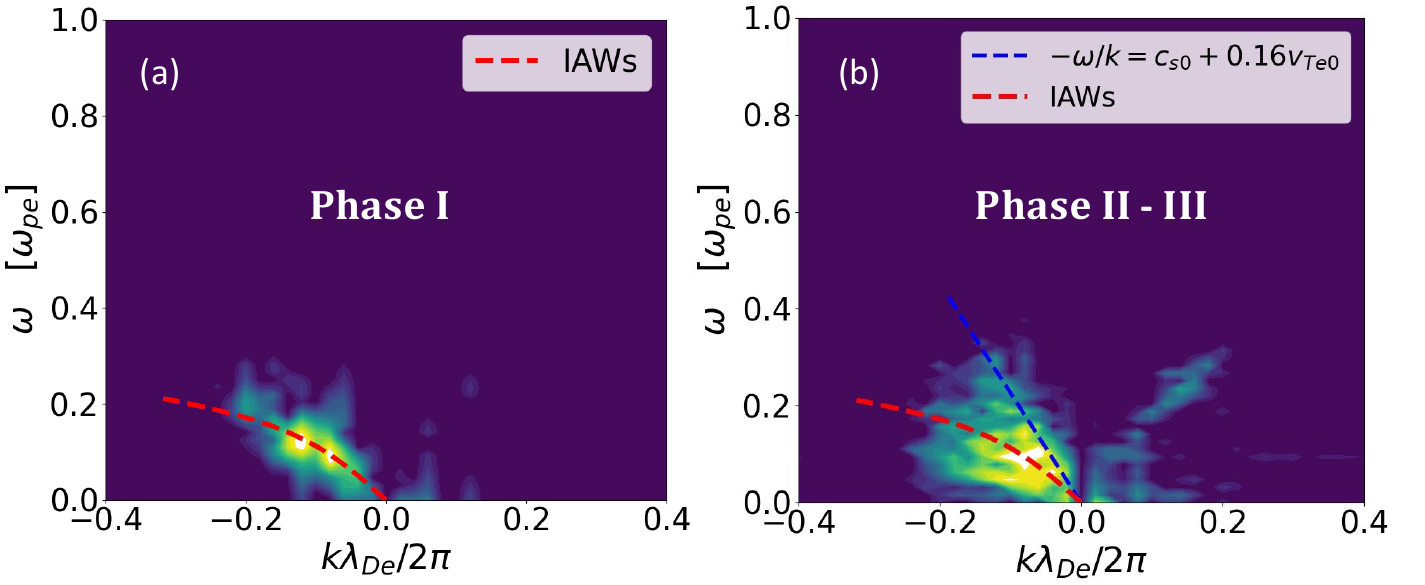}
\caption{\label{fig:5.omegak} Parallel wave energy spectrum $|E_z(\omega, k_z)|^2$ during the IAW burst phase (i.e., Phase I) (a), and the phase right after saturation (i.e., Phase II-III) (b) for Run {\tt M25E10}. The dispersion relation of IAWs (Eq.~\eqref{sup:eq:dieletric}) is plotted with the red dashed curve in (a) and (b). The blue dash line in (b) indicates a set of waves with positive frequency shifts in the nonlinear stage.}
\end{figure}

One signature of electron trapping is that it can trigger positive frequency shifts of nonlinear IAWs, which may allow the sub-harmonic decay of IAWs~\citep{berger2013electron,chapman2013kinetic,chapman2014new}. 
Modes with positive frequency shifts can spread across the region in $\omega-k$ space approximately bounded by $\omega/k = c_{s0} + \Delta v_i$, where $\Delta v_i$ corresponds to a region where ion Landau damping is reduced by ion trapping.
To obtain evidence of this, we examine the wave energy spectrum by performing 2D Fourier transforms (in the parallel direction ($z$) and in time) of the parallel electric field $E_z$ between different time ranges. 
Note that due to limited storage and computational power, we are not able to save the files often enough to have high resolution in frequency for Run {\tt Main}. 
Instead, the wave energy spectra $|E_z(k,\omega)|^2$ are calculated using Run {\tt M25E10} at the corresponding evolution phases.
In the linear phase, the $\omega-k$ diagram is consistent with the IAI linear dispersion relation, as shown in Figure~\ref{fig:5.omegak}(a).
In phases II-III, as shown in Figure~\ref{fig:5.omegak}(b), we indeed see the positive frequency shift of IAWs in a wide range of wavenumbers. 
The maximum shift of the phase speed, $\Delta v_i$, can be estimated using $\Delta v_e$, which is the half-width of the plateau of the electron velocity distribution before saturation (the plateau for Run {\tt Main} can be seen in Figure~\ref{fig:2.dist_1d} (a)), i.e., $\Delta v_i \approx \sqrt{m_e/m_i} \Delta v_e \approx 0.2 v_{Te0}$.
This estimate roughly agrees with the observed spectrum in Figure~\ref{fig:5.omegak}(b).

As mentioned above, this frequency shift is expected to induce sub-harmonic decay of IAWs and, consequently, modulate the energy spectrum. 
In~\citet{chapman2014new}, the onset of IAT leads to $\phi_k \propto k^{-4}$ at  relatively large wavenumbers.
The corresponding wave energy spectrum is then $N(k) \propto |E_k|^2 \propto k^2 \phi_k^2 \propto k^{-6}$.
This spectrum is observed in our simulation at the late stage of phase III and in phase IV, as evidenced by the dash-dotted curves in Figure~\ref{fig:4.spectrum}(a).

\subsubsection{Kadomtsev-Petviashvili spectrum}
Another noteworthy nonlinear feature observed in the wavenumber spectrum, as shown by the dashed curve in Figure~\ref{fig:4.spectrum}(a), is its agreement with the Kadomtsev-Petviashvili (KP) spectrum, characterized by $N(k) \sim 1/k^3 \ln(1/k\lambda_{De})$, immediately after the saturation stage ($\omega_{pe}t=1000$). 
While this spectrum has been observed in certain turbulent heating experiments~\citep{hamberger1972experimental,perepelkin1973nonlinear,de1991experimental}, we believe this to be its first direct numerical validation.

The agreement between our simulation results and the KP spectrum may initially seem puzzling, as the original derivation of the KP spectrum necessitates strong ion nonlinear effects and negligible ion Landau damping~\citep{kadomtsev1962turbulence,petviashvili1963ion}.
Although we confirmed earlier in section~\ref{sec:phase3-1} that Landau damping is indeed negligible right after saturation, the requirement for strong ion nonlinear effects still seems to contradict the quasi-linear saturation mechanism that we previously confirmed.
In fact, even though quasi-linear effects of electron distribution dominate around saturation, ion nonlinear effects (induced scattering by ions) can still become noticeable afterward, as evidenced by the emerging plateau in the ion distribution function around $\omega_{pe}t = 1000$ (see Figure~\ref{fig:2.dist_1d} (b)).
\citet{bychenkov1982ion} also demonstrated that the KP spectrum can be obtained by simultaneously considering the quasi-linear relaxation of the electron distribution, leading to a weakening of $\gamma_e$, and the induced scattering process, leading to a damping rate of $\gamma_{\rm{NL}}$.
Specifically, they showed that the condition $\gamma(\mathbf{k}) = \gamma_e (\mathbf{k}) + \gamma_i (\mathbf{k}) + \gamma_{\rm{NL}} (\mathbf{k}) = 0$ can lead to the emergence of the KP spectrum, even when $|\gamma_{\rm{NL}}(\mathbf{k})|\ll \rm{min}\{|\gamma_{e}(\mathbf{k})|,|\gamma_{i}(\mathbf{k})|  \}$~\citep{bychenkov1982ion,bychenkov1988ion}.
Our observation confirms the validity of their theory during the times around saturation and emphasizes the significance of nonlinear effects in the subsequent order of weak turbulence theory in shaping the behavior of IAT, even if they are of relatively minor importance in the saturation itself.

\subsection{\label{sec:phase4}Phase IV: post-shutdown phase}
At the end of phase III, the current growth rate returns to a value that is close to the free acceleration rate, as shown by the orange dashed line in Figure~\ref{fig:1.current_and_energy}.
The ion heating rate also relaxes to a relatively low level (only about 25\% ion heating happens after phase III, see Figure~\ref{fig:3.temp}). 
After IAT is shut down, the system still undergoes a long evolution process before entering a new regime. 
We refer to this nonlinear evolution stage as phase IV.

\begin{figure}
\includegraphics[width=1.0\textwidth]{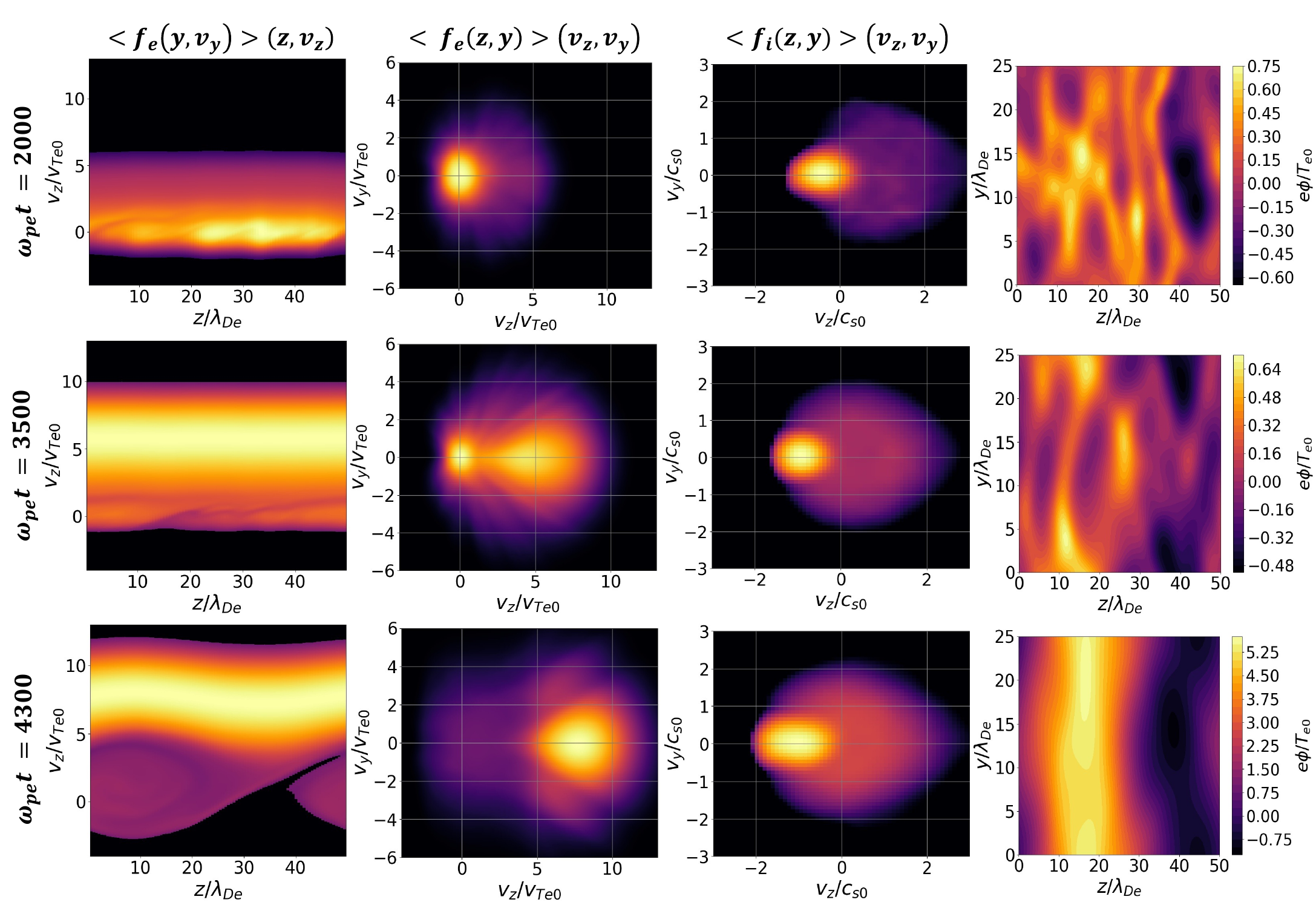}
\caption{\label{fig:6.2d} Snapshots of electron phase-space (first column), electron velocity distribution (second column), ion velocity distribution  (third column), and electric potential (last column) at the end of phase III, the end of phase IV, and the middle of phase V, respectively. The color bar for the ion velocity distribution is in logarithmic scale. The 2D distributions are obtained by averaging the 4-dimensional (2D2V) distribution over the other two dimensions. }
\end{figure}

To gain a better understanding of this stage, we plot the 2D features of the system in Figure~\ref{fig:6.2d}. 
Snapshots of the 2D electron velocity distribution are shown in the second column. 
At $\omega_{pe} t = 2000$, the bulk of the electron distribution is centered around the resonance region.
The subsequent shutdown of the IAI decreases the effective friction on the electrons, and allows the bulk to migrate to the tail region by $\omega_{pe} t = 3500$.
By the end of phase IV, the electron velocity distribution is greatly broadened in the parallel direction with respect to what it was at the end of phase III. 
Furthermore, we observe that the tail of the electron distribution exhibits a triangular shape, consistent with early analytical conjectures~\citep{sagdeev1969nonlinear}.
This is because oblique unstable modes have a maximum propagation angle with respect to the parallel direction. The electron distribution population that lies outside of the resonance region created by the wave modes is more easily accelerated by the external electric field.

\subsubsection{Formation of high-energy ion tail and dominance of oblique modes}
Another interesting feature is that a much more obvious high energy ion-tail forms, which becomes visible in the third column of Figure~\ref{fig:6.2d} at $\omega_{pe}t=2000$ and can be clearly seen at $\omega_{pe}t=3500$ (note that the colormap of the ion velocity distribution plots is in logarithmic scale to better emphasize its features).
By comparing the cyan, green, and yellow curves in Figure~\ref{fig:2.dist_1d}(b), we can observe that the cold bulk ions experience some heating during phase II and the early stages of phase III. However, the formation of the high-temperature tail primarily occurs during phase III and phase IV.
This phenomenon is consistent with a previous 2D simulation~\citep{dum1974turbulent} and can be explained by the quasi-linear theory for ions~\citep{ishihara1981quasilinear}.
Initially, as the ions are extremely cold, only a very small fraction of waves with low phase velocities ($v_{ph}<c_s$) can directly heat the cold bulk ions through weak ion trapping. 
However, as time progresses, the quasi-linear effect gradually diffuses the non-resonant portion of the ion distribution towards the phase-velocity range of the IAWs, allowing for stronger direct resonant interactions with the waves.
This quasi-linear diffusion of non-resonant ions is the primary mechanism contributing to the formation of the high-energy ion tail~\citep{ishihara1981quasilinear,ishihara1983quasilinear}. 
It is interesting to note that similar ion distribution is also observed in a recent 2D2V Vlasov simulation of current-driven instabilities~\citep{chan2022spectral}.

The ion tail formation can also help explain the fact that the most intense wave modes propagate with a non-zero angle with respect to the parallel direction during phase IV, as evidenced by the yellow curve in Figure~\ref{fig:4.spectrum} (b).
According to linear theory, parallel modes are expected to dominate over oblique modes because of their higher linear growth rates and saturation levels (because parallel modes have higher effective drift velocity).
This is confirmed by our simulations in the linear stage (phase I, the dark blue curve in Figure~\ref{fig:4.spectrum}(b)) and early times during the nonlinear evolution (phase III, the blue curve).
However, as the quasi-linear diffusion takes place, more ions are diffused in the parallel direction from the bulk (non-resonant portion) to the resonant region, which is confirmed by the third column of Figure~\ref{fig:6.2d} at $\omega_{pe}t=2000$ and $\omega_{pe}t=3500$.
As a result, modes with small propagation angles are more heavily damped because there are more ions participating in the resonant interactions with the waves.
Indeed, as shown by the yellow curve in Figure~\ref{fig:4.spectrum}(b), we see oblique modes dominating over parallel modes at $\omega_{pe}t=3500$.
This observation also clearly demonstrates the relevance of oblique modes in the nonlinear evolution of IAI, even in the weak electric field regime that we consider here.
At the later time of phase IV, the damping of the waves is gradually reduced by the formation of the plateau in the high-energy ion tail visible in Figure~\ref{fig:2.dist_1d}(b), which is confirmed by the red curve in Figure~\ref{fig:1.current_and_energy}.

\subsubsection{Formation of double layer}
Finally, we look at the electron phase space structures in the first column of Figure~\ref{fig:6.2d} and the corresponding electric potential in the last column.
In section~\ref{sec:phase3}, we mentioned the emergence of electron holes right after saturation in section~\ref{sec:phase3-1}.
During the later course of the nonlinear evolution, these phase-space vortexes gradually merge and finally become one single large electron hole at the end of phase IV (which can be seen from the electron phase space plot at $\omega_{pe}t = 3500$). 
In the meantime, the potential wells created by the waves become asymmetric (which means there is a non-vanishing electric potential change across a single potential well), with a steeper edge.
The steep asymmetric potential wells favor the triggering of new instabilities because they asymmetrically reflect low-energy electrons and accelerate high-energy electrons, which enhances the two-stream structure in the electron distribution and depletes the particles in the middle of the well.
In the electric potential depicted in the second row of Figure~\ref{fig:6.2d} at $\omega_{pe}t=3500$, we see clearly larger and more asymmetric potential structures compared with the electric potential at $\omega_{pe}t=2000$.
The impacts on the electron velocity distribution function can be seen in Figure~\ref{fig:2.dist_1d}(a), where the depletion of the electrons at around $v_{Te0}$ is already visible at $\omega_{pe}t =3500$.
Another possible consequence of particle trapping is modulational instability. In~\citet{rose2005langmuir}, it was demonstrated that negative nonlinear frequency shift and wave diffraction of Langmuir waves can trigger a self-focusing effect.
It is possible that a similar effect could extend to the IAWs.
In Figure~\ref{fig:7.omegak}(a), we indeed observe a stronger negative frequency shift during phase IV, which emphasizes the effects of nonlinear ion trapping at the late stage of the nonlinear evolution~\citep{berger2013electron}. 
The modulated wavefronts (the localized maxima in electric potential) observed in the last column of Figure~\ref{fig:6.2d} at $\omega_{pe}t=3500$, may result from this self-focusing effect.

Extending from phase IV to phase V, the potential wells coalesce and steepen further, and, finally, a double layer forms --- clearly visible in the rightmost column of Figure~\ref{fig:6.2d} at $\omega_{pe}t = 4300$.
The formation of double layers associated with IAI has been shown via experiments ~\citep[e.g.,][]{okuda1982ion,chanteur1983formation}, simulations~\citep[e.g.,][]{chanteur1985vlasov,sato1980ion}, and spatial measurements~\citep[e.g.,][]{ergun2001direct,ergun2009observations}. We here confirm the production of the double layer by IAT with a 2D2V simulation with a more self-consistent setup.

\subsection{\label{sec:phase5}Phase V: onset of electron-acoustic waves} 

\begin{figure}
\includegraphics[width=1.0\textwidth]{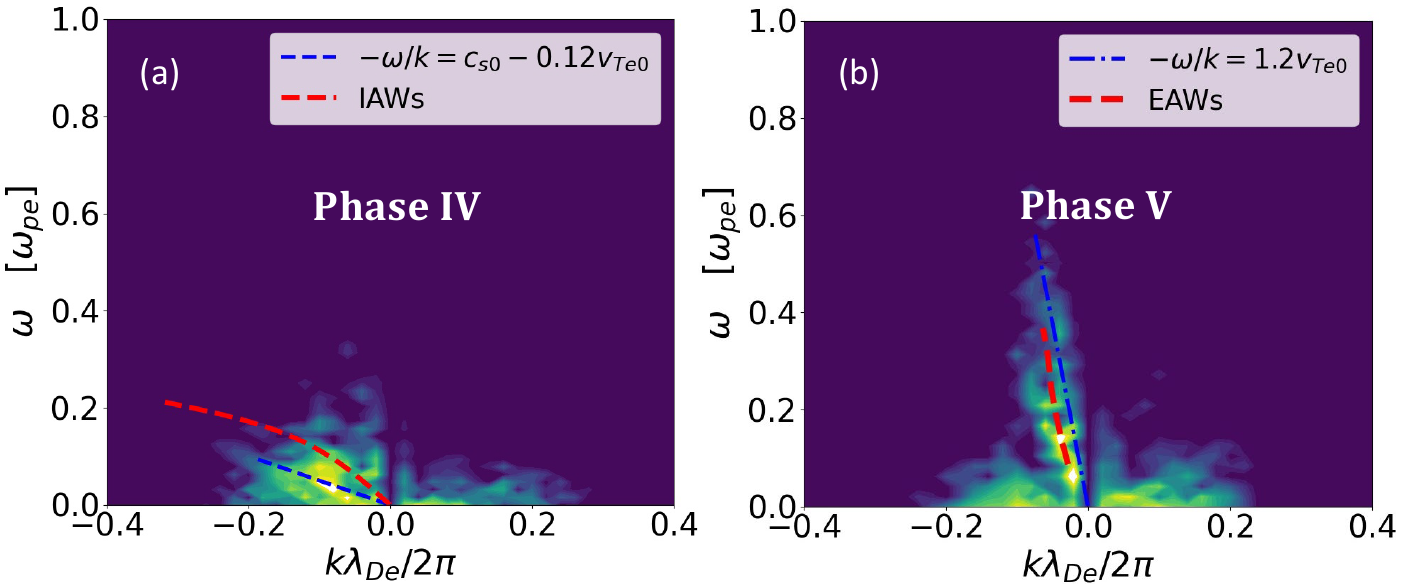}
\caption{\label{fig:7.omegak} Parallel wave energy spectrum $|E_z(\omega, k_z)|^2$ after effective shutdown of IAT (i.e., Phase IV) (a), and the EAW burst phase (i.e., Phase V) (b) for Run {\tt M25E10}. The dispersion relations of IAWs and EAWs are plotted with the red dashed curve in (a) and (b), respectively. The blue dash-dotted line in (a) indicates a set of waves with negative frequency shifts. The blue dash line in (b) indicates a set of waves with a phase velocity of $1.2v_{Te0}$.}
\end{figure}

As evidenced by the wave energy time trace in Figure~\ref{fig:1.current_and_energy}, another instability is triggered at this moment.
We interpret these new wave modes as electron-acoustic waves (EAWs)~\citep{holloway1991undamped, valentini2006excitation, valentini2012undamped,anderegg2009electron} for the reasons described below.

First, we examine the wave energy spectrum with Run {\tt M25E10} again, which is shown in Figure~\ref{fig:7.omegak}(b).
We can see that the new wave modes in phase V exhibit much higher frequencies extending from $0.1 \omega_{pe}$ to around $ 0.6 \omega_{pe}$ and a much faster phase velocity ($\sim 1.2v_{Te0}$), which excludes the possibility of (slow) ion modes.
Instead, the intermediate frequency range between ion and electron plasma frequencies is standard for EAWs~\citep{holloway1991undamped}.
Second, as seen from the electron phase structure at $\omega_{pe} t = 4300$ in the first column of Figure~\ref{fig:6.2d}, the electron distribution has a trapped population around $v_{Te0}$ with a trapping width extending to $\sim 4 v_{Te0}$.
These trapped electrons around $v_{Te0}$ effectively create a plateau shape around the phase velocity of the waves (i.e., $\sim v_{Te0}$) in the electron velocity distribution (which can be seen from the second column in Figure~\ref{fig:6.2d} at $\omega_{pe}t=4300$) and allow the EAWs to stay immune to Landau damping~\citep{holloway1991undamped}.
Third, we plot the theoretically obtained dispersion relation of EAWs by approximating the electron velocity distribution with two Maxwellian functions and a plateau in Figure~\ref{fig:7.omegak}(b), which agrees reasonably well with the wavenumbers and frequencies we observe in the simulation.
These EAWs have lower wavenumbers compared to IAWs, which is consistent with the domain-size electron-hole observed in the last row in Figure
~\ref{fig:6.2d}.
In appendix~\ref{app:3}, we describe in detail how we obtain the dispersion relation of EAWs.
The formation of double layers and bursts of EAWs after bursts of IAWs are also reported in recent experimental~\citep{zhang2022} and numerical~\citep{zhang2022, chen2022electron, hara2019ion, vazsonyi2020non} studies. 

\section{\label{sec:results_2} Dependence of results on simulation parameters}

In this section, we investigate the dependence of the results discussed above on mass ratio, external electric field, and initial temperature ratio.

\subsection{\label{sec:resistivity_strength} Anomalous resistivity intensity}

\begin{figure}
\includegraphics[width=1.0\textwidth]{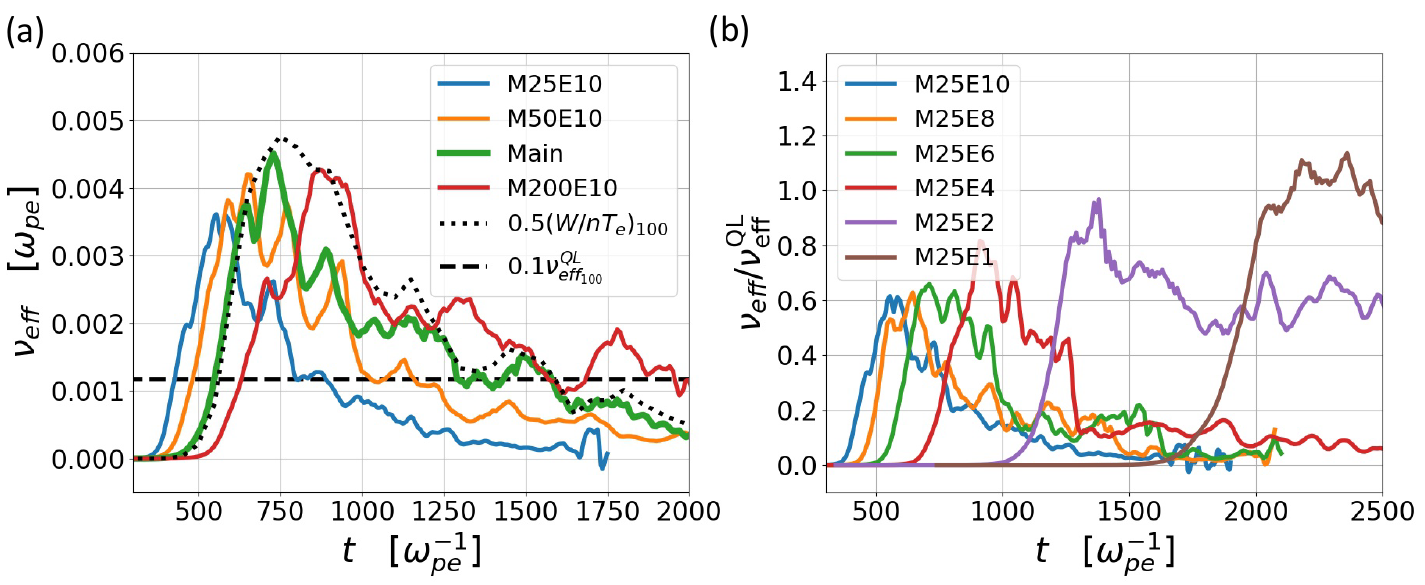}
\caption{\label{fig:8.nu_eff} (a) Time traces of effective collision frequency for simulations with different mass ratios and the same external electric field ($\tilde{E}_{\rm{ext}} = 2.5\times 10^{-3}$). 
For Run {\tt Main} we also plot, as a reference, the curve corresponding to $0.5 W/n_0T_e$, where $W$ is the wave energy density (dotted curve), and 10\% of the quasi-linear value (Eq.~\eqref{eq:nu_eff-QL}) (dashed horizontal line).
(b)  Time traces of effective collision frequency normalized to the quasi-linear prediction, Eq.~\eqref{eq:nu_eff-QL}, for simulations with different external electric fields and the same mass ratio ($m_i/m_e = 25$).
 }
\end{figure}

We define the effective collision frequency, $\nu_{\rm{eff}}$, from the equation $ dJ_z/dt = e^2n_0/m_e E_{\rm{ext}} - \nu_{\rm{eff}}J_z$.
A reference value is the quasi-linear result~\citep{bychenkov1988ion}:
\begin{align}
\label{eq:nu_eff-QL}
\nu_{\rm{eff}}^{\rm{QL}} \approx 0.47 \omega_{pe}\tilde{E}_{\rm{ext}}\left(\frac{m_i}{m_e}\right)^{1/2} .
\end{align}

Shown in Figure~\ref{fig:8.nu_eff}(a) are time traces of $\nu_{\rm{eff}}$ for runs with the same external electric field but different mass ratios. 
It can be seen that the observed $\nu_{\rm{eff}}$ in phase III is only about $10\% \sim 20\%$ of the value predicted by Eq.~\eqref{eq:nu_eff-QL} for Run {\tt Main}.
The maximum $\nu_{\rm{eff}}$, which is observed during saturation (in phase II), is also smaller than the quasi-linear value (about 38\% of  Eq.~\eqref{eq:nu_eff-QL}).
In terms of the mass ratio dependence, although the peak values depend weakly on the mass ratio, $\nu_{\rm{eff}}$ after saturation shows a roughly linear dependence on $\sqrt{m_i/m_e}$. 
For example, Run {\tt Main} (green curve) exhibits approximately twice the amplitude of Run {\tt M25E10} (blue curve) after saturation.

The main disagreement between the simulation results and the quasi-linear theory is that the latter assumes the electron-ion drift velocity will be fixed at around the ion sound speed after IAI saturates~\citep{rudakov1966quasilinear,bychenkov1988ion}. 
This assumption is only valid for the resonant part of the electrons.
However, both resonant and non-resonant parts of electrons are accounted for in the current.
Due to the fact that most of the tail, with a significantly higher drift velocity, does not resonate with the waves and is freely accelerated by the external electric field, the  $\nu_{\rm{eff}}$ we obtain in our simulations is much smaller than the theoretical one and continues to decrease as the current grows, even if the friction force on the bulk electrons provided by the scattering of waves is unchanged.

Figure~\ref{fig:8.nu_eff}(b) shows the time traces of $\nu_{\rm{eff}}/\nu_{\rm{eff}}^{\rm{QL}}$ for simulations with the same mass ratio ($m_i/m_e=25$) but different external electric fields.
As long as the electric field is relatively strong (but still in the weak field regime, i.e. $E_{\rm{ext}} \ll E_{\rm{NL}}$), the linear dependence of $\nu_{\rm{eff}}$ on the external electric field holds.
For example, $\nu_{\rm{eff}}$ in Run {\tt M25E10}, Run {\tt M25E8}, and Run {\tt M25E6} have nearly the same normalized amplitudes.
As the electric field becomes weaker, the effective collision frequency starts to approach the quasi-linear value during saturation gradually.
For Run {\tt M25E1}, the effective collision frequency roughly agrees with the quasi-linear prediction around saturation.

Weak enough external electric fields such that Eq.~\eqref{eq:nu_eff-QL} approximately holds will be referred to as {\it extremely weak}. 
In this {\it extremely weak field regime},  electrons have a drift velocity very close to the ion sound speed (i.e., the phase velocity of the IAW) at the moment of saturation and, therefore, most of the electron population, including the tail, can be trapped in the resonance region after saturation, which validates the assumptions underlying quasi-linear theory. 
Because the tail portion cannot effectively run away in this case, the electron distribution barely changes during phase III. 
The wave energy also remains approximately steady, pinned at the saturation level.
Otherwise, most physics discussed in section~\ref{sec:results_1} still holds, including the saturation and shutdown mechanisms, and transition into bursts of EAWs.
An estimate of the threshold electric field and more details on the {\it extremely weak field regime} can be found in appendix~\ref{app:4}.

Another well-known formula for anomalous resistivity, due to Sagdeev~\citep{Sagdeev1967, zavoiski1967turbulant},
\begin{align}
\label{eq:nu_eff-Sagdeev}
\nu_{\rm{eff}}^{\rm{Sagdeev}} \approx 0.2\omega_{pe}\left(\frac{T_e}{T_i}\right)^{1/2} \left( \frac{T_{e0}}{2T_e} \tilde{E}_{\rm{ext}}^2\right )^{1/4},
\end{align}
predicts an effective collision frequency of $\sim 0.06\omega_{pe}$ for Run {\tt Main}, which is over an order of magnitude larger than observed.
Such discrepancy is unsurprising because the simulations are performed in the weak electric field regime, whereas Sagdeev's formula is only expected to apply when $E_{\rm{ext}}>E_{\rm{NL}}$.

Finally, to compare our simulations with previous numerical studies~\citep{Labelle1988, watt2002ion, hellinger2004effective}, it is convenient to calculate a more qualitative estimate, namely, 
\begin{equation}
\label{eq:nu_eff-estimate}
    \nu_{\rm{eff}} \sim \omega_{pe} \frac{W}{n_0T_e},
\end{equation}
with $W$ being the wave energy density. 
This estimate is obtained by balancing the electron momentum loss due to the emission of waves and momentum gain from the external electric field. 
In fact, Sagdeev's formula also originates from this estimate, but it severely overestimates the wave energy by assuming that the nonlinear scattering of ions is the only saturation mechanism.
Eq.~\eqref{eq:nu_eff-estimate} for Run {\tt Main} is plotted as the black dotted curve in Figure~\ref{fig:8.nu_eff}(a), and turns out to be a reasonable match, especially after saturation (phases III \& IV).
This conclusion is at odds with values of anomalous resistivity exceeding this estimate by at least one order of magnitude reported in many previous numerical studies~\citep[e.g.,][]{watt2002ion, hellinger2004effective, petkaki2003anomalous,petkaki2006anomalous}.
Again, the fact that most previous simulations start with a super-critical configuration can lead to artificially high wave energy, which might be responsible for the large discrepancy between previous numerical experiments and Eq.~\eqref{eq:nu_eff-estimate}.

To summarize, no nonlinear theory that we are aware of gives a reliable estimate of the anomalous resistivity created by IAT in the weak electric field regime.
When the electric field is extremely weak, the quasi-linear value, Eq.~\eqref{eq:nu_eff-QL}, is a good approximation. 
Otherwise, the anomalous resistivity after saturation is about $10\%\sim 20\%$ of Eq.~\eqref{eq:nu_eff-QL} in our simulations.
We also find that Eq.~\eqref{eq:nu_eff-estimate} provides a reasonable fit.
However, even though we have validated Eq.~\eqref{eq:W_sat} for the wave energy at saturation, the fact that $W$ does not remain at that level, but rather immediately starts to decay, precludes \textit{a priori} usage of this estimate.

\subsection{\label{sec:resistivity_length} Particle heating and anomalous resistivity duration}
As explained in section~\ref{sec:phase3}, anomalous resistivity will eventually become negligible as a result of the effective shutdown of the IAT.
Therefore, in addition to characterizing the intensity of the anomalous resistivity, it is of interest to understand how long it is sustained at appreciable levels as a function of mass ratio, external electric field, and initial temperature ratio.
Because the shutdown of IAT is mainly associated with temperatures, we investigate how electron and ion heating depends on these parameters. 

Before we proceed with our analysis, we stress again that we define temperature as the second moment of the distribution function.
This deviates from temperature in the normal sense, because neither the electron's nor the ion's distribution functions are Maxwellian when the system enters the nonlinear stage.
This makes it challenging to derive an analytical expression that precisely describes the temporal evolution of electron and ion heating and the shutdown criterion of IAT.
\begin{figure}
\centering
\includegraphics[width=0.6\textwidth]{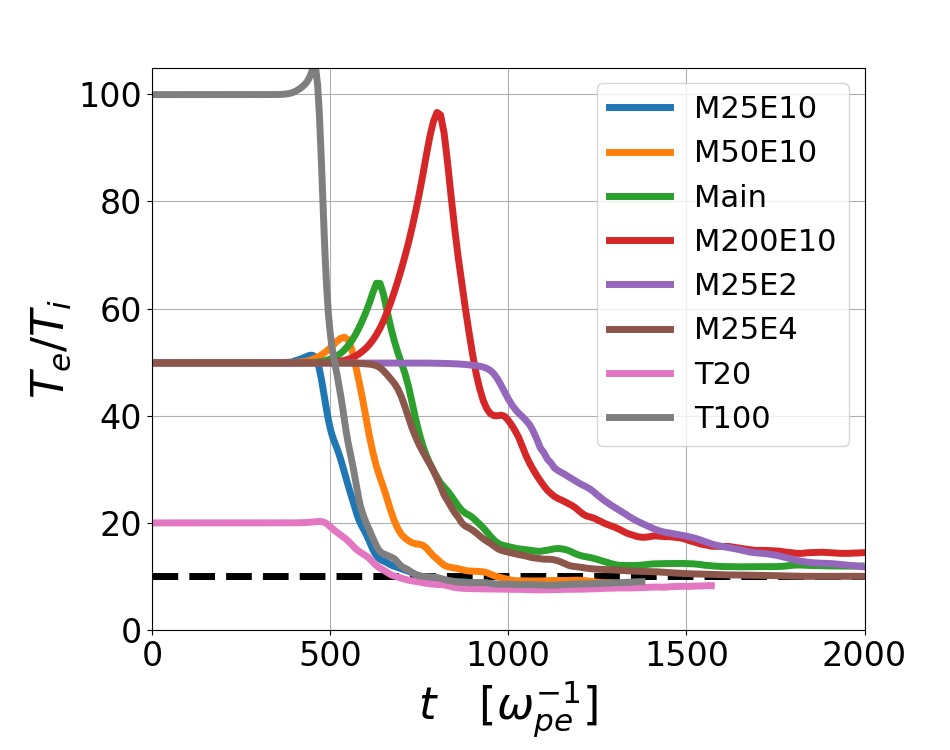}
\caption{\label{fig:9.temp_ratio} Time traces of temperature ratios for simulations with different mass ratios, electric fields, and initial temperature ratios. The black dashed line marks $T_e/T_i = 10$.}
\end{figure}

\begin{figure}
\centering
\includegraphics[width=0.6\textwidth]{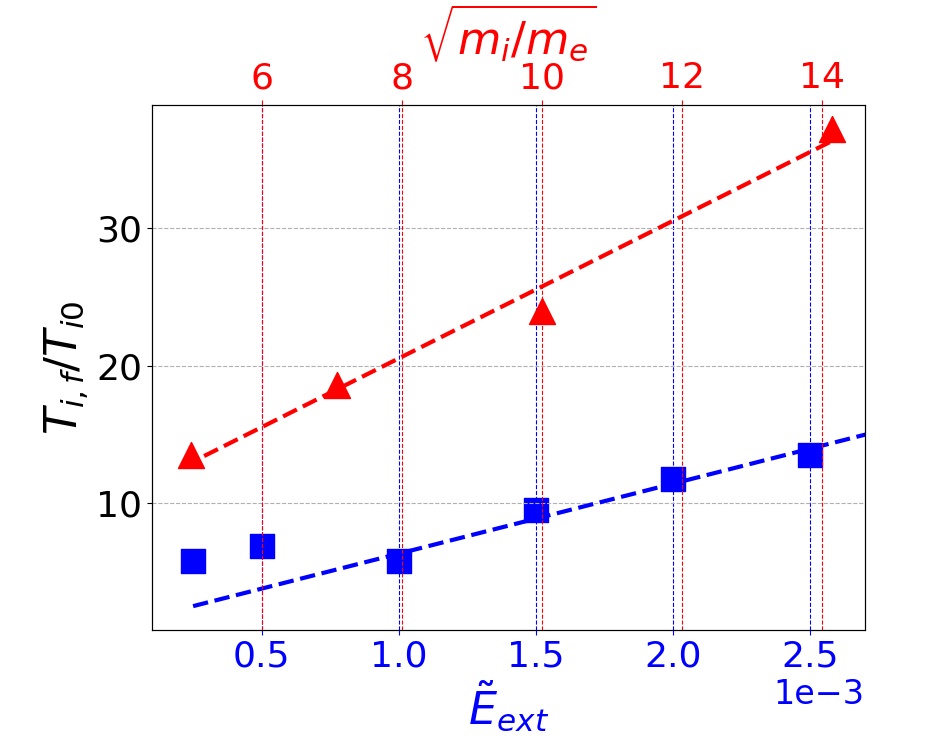}
\caption{\label{fig:10.ti} Final ion temperature (normalized to its initial value) as a function of the square root of mass ratio at fixed $\tilde{E}_{\rm{ext}}=2.5 \times 10^{-3}$ (top horizontal axis, red) and as a function of the external electric field at fixed  $m_i/m_e= 25$ (bottom horizontal axis, blue).}
\end{figure}

\begin{figure}
\centering
\includegraphics[width=0.6\textwidth]{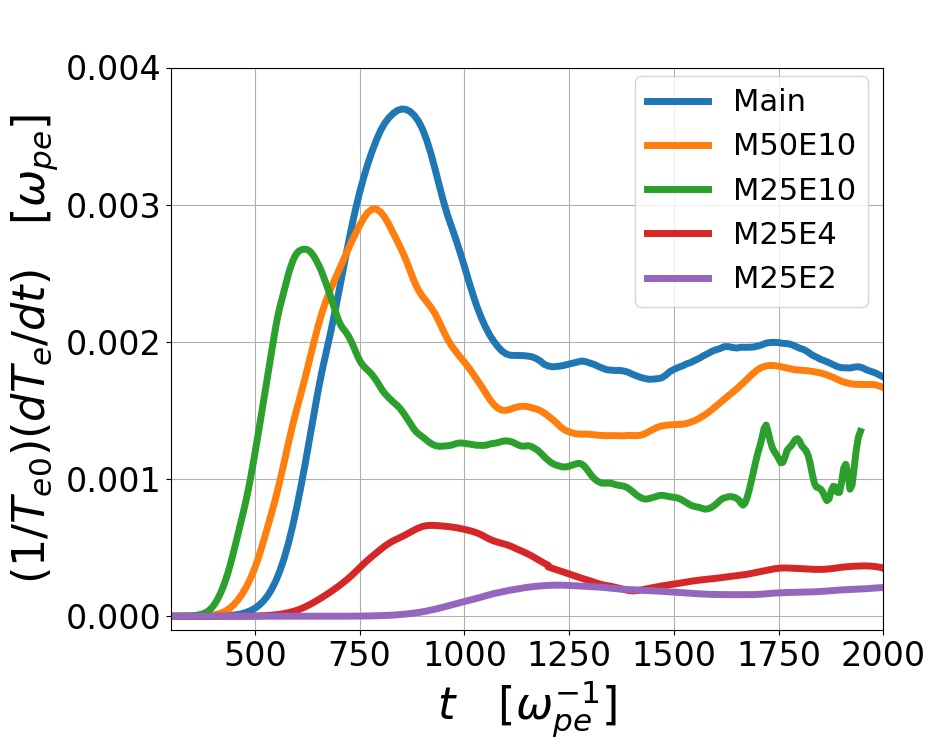}
\caption{\label{fig:11.te} Time traces of electron heating rate for some example simulations with different mass ratios and electric fields. }
\end{figure}

\subsubsection{Ion and electron heating}
The time traces of electron-to-ion temperature ratio for simulations with different mass ratios, external electric fields, and initial temperature ratios are shown in Figure~\ref{fig:9.temp_ratio}.
It can be seen that all our simulations reach a final temperature ratio of $\sim 10$.
Moreover, as shown by {\tt T20} (pink curve) and {\tt T100} (silver curve) in Figure~\ref{fig:9.temp_ratio}, the fact that the final temperature ratio (compared with the blue curve) is independent of the initial temperature ratio confirms the conjecture that IAT is indeed shut down by reaching a specific temperature ratio. 

For use in what follows, let us define the final temperature ratio as
\begin{equation}
    \label{eq:shutdown_criterion}
    \frac{T_{e,\rm{f}}}{T_{i,\rm{f}}} = \theta,
\end{equation} 
where $T_{i,\rm{f}}$ and $T_{e,\rm{f}}$ denote the final ion and electron temperature, i.e. the temperature when the ion heating is effectively shut down.
The effective shutdown is defined as the moment when the ion heating rate reduces to 10\% of its peak value.
Note that $T_{i,\rm{f}}$ is a little bit larger than the ion temperature at the end of phase III because there is still some ion heating happening during phase IV.
Figure~\ref{fig:10.ti} plots $T_{i,\rm{f}}$ for simulations with different mass ratios and external electric fields, showing a linear dependence on $\sqrt{m_i/m_e}$ and $E_{\rm{ext}}$.
This is the same as the dependence of $W_{\rm{sat}}$ on $E_{\rm{ext}}$ and $\sqrt{m_i/m_e}$ in Eq.~\eqref{eq:W_sat} and confirmed in table~\ref{tab:saturation}.
We thus fit the ion temperatures in most of our simulations as
\begin{equation}
    \label{eq:ti}
   n_0 T_{i,\rm{f}}  \approx 28.6 W_{\rm sat} .
\end{equation}
According to Eq.~\eqref{eq:shutdown_criterion} and Eq.~\eqref{eq:W_sat}, the electron temperature at the end of phase III can then be fitted by
\begin{equation}
\label{eq:te}
\begin{split}
n_0T_{e,\rm{f}} &\approx 28.6 \theta W_{\rm sat} \\
 &\approx 20.2 \theta\left(\frac{m_i}{m_e}\right)^{1/2} \tilde{E}_{\rm{ext}} n_0T_{e0}.
\end{split}
\end{equation}
However, there are two simulations that do not conform to Eq.~\eqref{eq:ti} in Figure~\ref{fig:10.ti} --- Run {\tt M25E1} and Run {\tt M25E2} (the leftmost two blue rectangular data points).
As discussed in section~\ref{sec:resistivity_strength}, these two simulations are under the {\it extremely weak field regime}.
In fact, the final ion temperature $T_{i, \rm{f}}$ given by Eq.~\eqref{eq:ti} can potentially become insufficiently low to satisfy the shutdown criterion, as described in Eq.~\eqref{eq:shutdown_criterion}. This scenario arises particularly when the external electric field is extremely weak.
This implies the existence of another threshold for the external electric field below $E_{\rm NL}$ --- which we call the {\it extremely weak field regime}.
Under this regime, as mentioned above, wave energy will be sustained at a high level during phase III. As the ion heating rate is proportional to the wave energy, being in this regime results in more intense ion heating than the prediction given by Eq.~\eqref{eq:ti} until the shutdown criterion Eq.~\eqref{eq:shutdown_criterion} is finally met.

\subsubsection{Duration of significant anomalous resistivity}
Finally, to estimate the duration of phase III, we plot time traces of electron heating rates for simulations with different mass ratios and external electric fields in Figure~\ref{fig:11.te}.
Both the time-dependent patterns and amplitudes of the electron heating rates are very similar to the effective collision frequencies shown in Figure~\ref{fig:8.nu_eff}(a), especially during phase II and the early stage of phase III, which is expected because electron heating is caused by anomalous resistivity.
However, unlike the anomalous resistivity, the electron heating rate does not die away but remains flat at later times in phase III.
This is due to the definition of temperature that we have adopted --- the increase in electron temperature is in fact mainly caused by the acceleration of the electron tail during phase III (so the overall electron distribution becomes broader). 
To provide a rough quantitative estimate of the duration of significance anomalous resistivity, we may approximate the electron heating rate with a constant value according to Figure~\ref{fig:11.te} as $ \left(1/T_{e0} \right) dT_e/dt \approx 0.08 \tilde{E}_{\rm{ext}}  \left(m_i/m_e\right)^{1/2} \omega_{pe}$, which has the same dependence on external electric field and mass ratio as the Eq.~\eqref{eq:nu_eff-QL}.
Then we can use $\tau_{\rm{res}} \approx \left(T_{e,\rm{f}} - T_{e0}\right) / (d T_e /dt )$ and substitute $T_{e,\rm f}$ with Eq.~\eqref{eq:te}. We arrive at
\begin{equation}
\label{eq:duration}
    \tau_{\rm{res}} \approx \frac{20.2 \theta (m_i/m_e)^{1/2}\tilde{E}_{\rm{ext}} -1 }{0.08 (m_i/m_e)^{1/2}\tilde{E}_{\rm{ext}}} \omega_{pe}^{-1}.
\end{equation}
While the exact value of $\theta$ depends on parameters, our simulations suggest that $\theta \approx 10$ (it is slightly larger for cases with a larger mass ratio and a smaller external electric field, see Figure~\ref{fig:9.temp_ratio}). 
This is consistent with linear theory~\citep{papadopoulos1977review,benz2012plasma}, and is justified by the fact that IAI becomes stable when $T_e/T_i<10$ (as long as the drift velocity between electrons and ions is much smaller than electron thermal velocity).
However, it is important to note that the linear theory is based on Maxwellian distribution functions, while in our simulations both the electron and ion distributions are rather better approximated by double-Maxwellians.
Therefore, using temperature as the sole parameter cannot capture the exact dynamics and the corresponding switch-off criterion accurately. This may explain the deviations of $\theta$ from the value $10$. 
Note that Eq.~\eqref{eq:duration} does not apply to {\it extremely weak field regime} due to the underestimate of electron and ion heating.

With Eq.~\eqref{eq:duration}, we are able to estimate the extent of the time interval over which the IAT-induced anomalous resistivity will last in a particular system.
For our Run {\tt Main}, Eq.~\eqref{eq:duration} yields $\tau_{res} \sim 2000 \omega_{pe}^{-1}$ if we adopt $\theta = 10$, which is in reasonable agreement with the numerical data shown in Figure~\ref{fig:8.nu_eff}(a).

\section{\label{sec:conclusion}Conclusion and discussion}

In this study, we present a comprehensive first principle (i.e., continuum Vlasov-Poisson) numerical investigation of the current-driven IAI in unmagnetized plasmas. 
By starting from a stable plasma equilibrium and gradually driving it towards instability using a (weak) external electric field, we ensure the self-consistent evolution of the system towards instability and the physical realizability of the unstable states that are reached.

We identify five distinct phases in the evolution of IAI and focus on the nonlinear phases. 
After the linear stage (phase I) --- the onset of which occurs when the linear growth rate of the instability approximately matches the current growth rate --- we observe (phase II) that the saturation of IAI primarily occurs through the quasi-linear relaxation of the electron velocity distribution.
We also observe that the 2D (in position space) system is much more efficient than its 1D counterpart in scattering electrons to the resonance region with lower drift velocity, which helps to form a resonant bulk and a tail in the electron velocity distribution.
Moving into phase III, contrary to the assumption made in many nonlinear or quasi-linear theories that a steady state is reached,  we find that the current continues to grow, albeit at a significantly reduced rate. 
This ongoing growth is attributed to the non-resonant electrons, which can still be accelerated by the external electric field. 
Additionally, the wave energy decreases due to the intense ion heating, ultimately leading to the suppression of IAT and related phenomena, including anomalous resistivity and ion heating.
We also observe the Kadomtsev-Petviashvili (KP) spectrum around saturation, which had not been reproduced numerically ever before and emphasizes the role of the nonlinear effect in setting the wavenumber spectrum, even though the effect itself is not the main driver of the saturation.
The effect of electron trapping and sub-harmonic decay of IAWs is also identified to modulate the wave spectrum.
In phase IV, we observe that the current growth rate approximately returns to the rate of free electron acceleration, due to the absence of significant wave-induced friction. 
Furthermore, we note the emergence of a high-energy ion tail and the prevalence of oblique modes during this phase. 
Finally, at the conclusion of our simulation, we observe the formation of a double layer in the electric potential, which triggers the generation of EAWs.

In the second part of our study, we shift our focus to the investigation of anomalous resistivity resulting from IAT and its dependence on various simulation parameters. 
We study the dependence of electron and ion heating on those parameters and derive an estimate for the time duration of significant anomalous resistivity.
We find that the quasi-linear approach yields the correct dependence on the mass ratio and external electric field; however, it generally predicts a larger absolute value of resistivity compared to our simulations. 
This discrepancy contradicts the findings of many previous 1D numerical studies, where stronger anomalous resistivity is typically observed.
We also observe, and derive analytically, a new electric field threshold below which the runaway electron population is small and the quasi-linear prediction of anomalous resistivity holds. 

Our study of the classical version of the IAT can be directly applicable to some physical systems. 
For instance, IAI can exist in the separatrix region of reconnecting current sheets, where the magnetic field may be negligible (in the no guide-field case). 
Situations where electrons are heated to high temperatures while ions remain cold, such as stellar flares~\citep{polito2018broad} and the low-speed solar wind near the Sun~\citep{verscharen2022electron,mozer2022core}, have been observed.
The findings in this paper are therefore directly relevant to such environments.
To be more specific, our simulations reveal features that match with direct observations in various plasma environments. 
For example, the plateaued-shaped ion velocity distribution with a high-energy tail which we find in our simulations is qualitatively similar to those measured in the solar wind~\citep{Mozer2020a} and interplanetary shocks~\citep{wilson2020electron}, where IAWs are believed to be present. 
We also demonstrate the IAT's capability to significantly heat the electron core and potentially create the ``strahl'' in the electron distribution commonly observed in the solar wind~\citep{verscharen2019multi}.
The development of anisotropy in particle velocity distribution is qualitatively similar to the measured electron distribution on the dayside of reconnecting magnetopause during IAW bursts~\citep{uchino2017waves,steinvall2021large}. 
Additionally, bursts of EAWs following IAT are experimentally observed in laser-driven reconnection events~\citep{zhang2022}.
We also observe that oblique modes dominate over parallel modes in the later stages of nonlinear evolution. 
When the driving field is strong, the dominance of oblique modes is expected to occur even earlier~\citep{bychenkov1988ion}.
These findings align with recent IAW observations in the solar wind~\citep{mozer2021triggered}, where the most intense waves propagate obliquely. 
The broadening of the spectrum during the nonlinear stage, as reported in other studies~\citep{Mozer2020b}, is also consistent with the wavenumber spectrum observed in our simulations (see Figure~\ref{fig:4.spectrum} (a)).

Our results can also be adapted to study magnetic reconnection influenced by IAT.
The electric field associated with reconnection is $E_{\rm{rec}}\sim \gamma V_A B_0$, where $\gamma$ is the reconnection rate and $V_A$ is the Alfv\'en speed computed with the reconnecting field $B_0$. 
Situations where this electric field is smaller than $E_{\rm NL}$ (Eq.~(\ref{eq:E_NL})) can exist in space plasmas, such as locations of the solar wind near the Sun (at distances around 20 solar radii) where $T_e/T_i \sim 5$ and $T_e \sim 50 eV$~\citep{mozer2021triggered}. 
For these parameters, we find $E_{\rm{rec}}/E_{\rm{NL}} \sim 1.2\pi (1/\beta_e) (v_{Te}/c) \sqrt{m_i/m_e} (T_i/T_e) < 1$.
In this case, assuming a fast reconnection rate of $\gamma\approx 0.1$, the reconnecting electric field is about $E_{\rm rec}/(4\pi e n_0\lambda_{De}) \approx 2\gamma\beta_e  \sqrt{m_e/m_i} (v_{Te}/c) \approx 5\times10^{-5}$, which should be in the {\it extremely weak field regime} that we have identified in this paper.
We may estimate the upper limit of the duration of significant resistivity by assuming that ion heating caused by IAT is an order of magnitude larger than the value given by Eq.~\eqref{eq:ti} (ion heating in Run {\tt M25E2} is about twice as intense as the prediction given by Eq.~\eqref{eq:ti}).
Then, according to Eq.~\eqref{eq:duration}, the corresponding $\tau_{\rm res}$ is at most on the order of $10^3\omega_{pi}^{-1}$.
On the other hand, the typical reconnection time, $\tau_{\rm{rec}}=\gamma^{-1} L/V_A$, where $L$ is the characteristic length scale of the reconnecting field, can also be estimated in this context.
There is a range of potential values of $L$ that one may consider; a reasonable lower bound appears to be the scale at which the turbulent inertial range starts ---  around $10^3\rm{km}$ at distances of this order~\citep{coles1988solar, lotz2023radial,huang2023new}.
At such a scale, the corresponding $V_A$ can be estimated using the Alfv\'en velocity at the outer scale and the observed magnetic power spectrum; a value of $V_A\approx 100 \rm{km/s}$ seems reasonable~\citep{huang2023new}. 
Assuming a fast reconnection rate of $\gamma \approx 0.1$, $\tau_{\rm{rec}}$ can easily extend beyond $10^{5} \omega_{pi}^{-1}$.
Therefore, we conclude that there is ample time for IAT to induce particle heating and change the particle distribution during a reconnection event in this system (and note that this conclusion should remain valid even if our estimates of $L$ and $V_A$ are somewhat off).

Finally, we discuss how our choice of simulation domain size and boundary conditions may limit the validity of our results.
The employment of periodic boundary conditions, which effectively imitates an infinitely extended plasma, raises the question of whether the simulated system size corresponds to realistic plasma scales.
In Run {\tt Main}, we can estimate the travel distance of an electron by integrating the area under the blue curve in Figure~\ref{fig:1.current_and_energy}, which is approximately $10^{4} \lambda_{De}$.
Hence, our conclusions should remain applicable in a real plasma as long as the system size is larger than this estimate.
In fact, this length scale ($10^4\lambda_{De}$) is small compared to the system size in various plasmas.
Considering the solar wind as an example, the Debye length at 1 AU is on the order of $10 \ \rm{m}$ and, so,
$10^{4} \lambda_{De}$ corresponds to a distance $\sim 100 \rm{km}$, which is still smaller than the ion skin depth ($\sim 140 \rm{km}$)~\citep{verscharen2019multi}.
On the other hand,  the limitation imposed by the small size of the simulation box, especially in the parallel direction, excludes the effects of long-wavelength wave modes that become prominent in phase V.
The box-size double layer that we observe can result from this artificial constraint.
Despite this limitation, we believe that the simulations conducted up to phase V and the mechanism elucidated for triggering the second instability remain plausible.
Expanding the simulation size in the parallel direction for a 2D2V simulation, while desirable, was deemed impractical due to the large computational resources it would require.

In summary, our numerical investigation represents a significant stride toward a self-consistent understanding of the nonlinear dynamics of IAT. 
While we concentrate on the weak field regime in this step, which contains simpler (though nonetheless complex) nonlinear physics than the strong field case, our study yields substantial insights applicable to various physical scenarios where IAWs are present.
Our study lays a robust foundation for future investigations into more complex IAT situations. 
These encompass strong external fields, the presence of magnetic fields, and less extreme initial temperature ratios. 
It is also important to acknowledge that our simulations do not encompass the hypothesized ultimate state of the system, wherein the current finally ceases to increase. 
The mechanism responsible for stopping particle acceleration when a collisionless plasma is exposed to an external electric field thus remains unknown. 
To better understand this problem, longer and more computationally demanding simulations would be required, which are currently beyond our available resources.
\newline

\paragraph{\textbf{Acknowledgments.}}
The authors thank I. Hutchinson, and N.~Mandell for insightful discussions.
The authors also thank the Gkeyll team for their help with the simulations.
This work was supported by the NSF-DOE Partnership in Basic Plasma Science and Engineering Award No. PHY-2010136 (ZL and NFL) and the Partnership for Multiscale Gyrokinetic Turbulence (MGK), part of the U.S. Department of Energy (DOE) Scientific Discovery Through Advanced Computing (SciDAC) program, via DOE contract DE-AC02-09CH11466 for the Princeton Plasma Physics Laboratory and subaward No. UTA18-000276 to M.I.T. under U.S. DOE Contract DE-SC0018429 (MF).
This research used resources of the MIT-PSFC partition of the Engaging cluster at the MGHPCC facility, funded by DOE award No. DE-FG02-91-ER54109 and the National Energy Research Scientific Computing Center, a DOE Office of Science User Facility supported by the Office of Science of the U.S. Department of Energy under Contract No. DE-AC02-05CH11231 using NERSC award FES-ERCAP0020063.
\newline

\paragraph{\textbf{Data availability statement.}}
Readers may reproduce our results and also use Gkeyll for their applications. The code used in this study is available online. Full installation instructions for Gkeyll are provided on the Gkeyll website~\footnote{https://gkeyll.readthedocs.io.}.
The input files used in the study are under version control and can be obtained from the repository at \url{https://github.com/ammarhakim/gkyl-paper-inp/tree/master/2023_JPP_iat}.

\appendix

\section{Linear theory of ion-acoustic instability~\label{app:0}}
Starting from the linearized Vlasov-Poisson equations Eq.~\eqref{eq:vlasov-possion}, the plasma dielectric function can be written as~\citep{schekochihin2022lectures}
\begin{equation}
    \epsilon(p, k) =1-\sum_{\alpha} \frac{\omega_{\mathrm{p} \alpha}^{2}}{k^{2}} \frac{i}{n_{\alpha}} \int \mathrm{d}^{3} \boldsymbol{v} \frac{1}{p+i \boldsymbol{k} \cdot \boldsymbol{v}} \boldsymbol{k} \cdot \frac{\partial f_{0 \alpha}}{\partial \boldsymbol{v}},  
\end{equation}
where $p = -i\omega + \gamma$ and $\alpha$ refers to different species.
Given a wave-vector $\boldsymbol{k}$, one can always choose the $z$-axis to be along$\boldsymbol{k}$.
Therefore, the dielectric function can be rewritten as
\begin{equation}
        \epsilon(p, k) =1-\sum_{\alpha} \frac{\omega_{\mathrm{p} \alpha}^{2}}{k^{2}} \frac{i}{n_{\alpha}} \int \mathrm{d}v_z \frac{F_{\alpha}'(v_z)}{v_z-i p/k},
\end{equation}
where $F_\alpha(v_z)$ is the 1D velocity distribution function, and prime denotes differentiation with respect to $v_z$. The analytical dispersion relation can be obtained by solving 
\begin{equation}
\label{sup:eq:dieletric}
    \epsilon(p,k) = 0.
\end{equation}

\begin{figure}
  \centering
\includegraphics[width=1.0\textwidth]{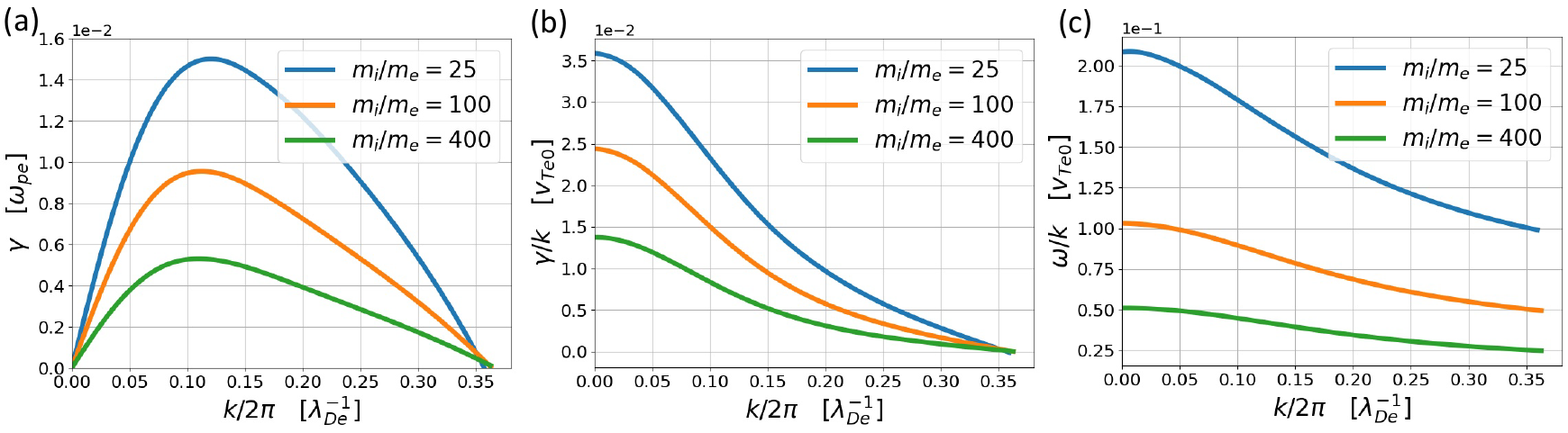}
\caption{Growth rates (a), growth rates divided by wavenumbers (b), and phase velocities (c) of IAWs for different mass ratio cases obtained from solving Eq.~\eqref{sup:eq:dieletric}.}
\label{sup:fig:linear_theory}
\end{figure}

Figure~\ref{sup:fig:linear_theory} shows plots of the growth rates ($\gamma$), growth rates divided by wavenumbers ($\gamma/k$), and phase space velocities ($\omega/k$) as functions of the wavenumber, obtained by solving Eq.~\eqref{sup:eq:dieletric} for cases with different mass ratios. 
Both electrons and ions have a Maxwellian velocity distribution. The electron distribution has a drift velocity of $0.5 v_{Te0}$ with respect to ions.
The electron-to-ion temperature ratio is 50, as in most of our runs. 
Figures~\ref{sup:fig:linear_theory}(b) and ~\ref{sup:fig:linear_theory}(c) are useful for determining the numerical resolution necessary; see appendix~\ref{app:5}.

\section{Comparison with one-dimensional simulation~\label{app:1}}

\begin{figure}
  \centering
  \includegraphics[width=0.9\textwidth]{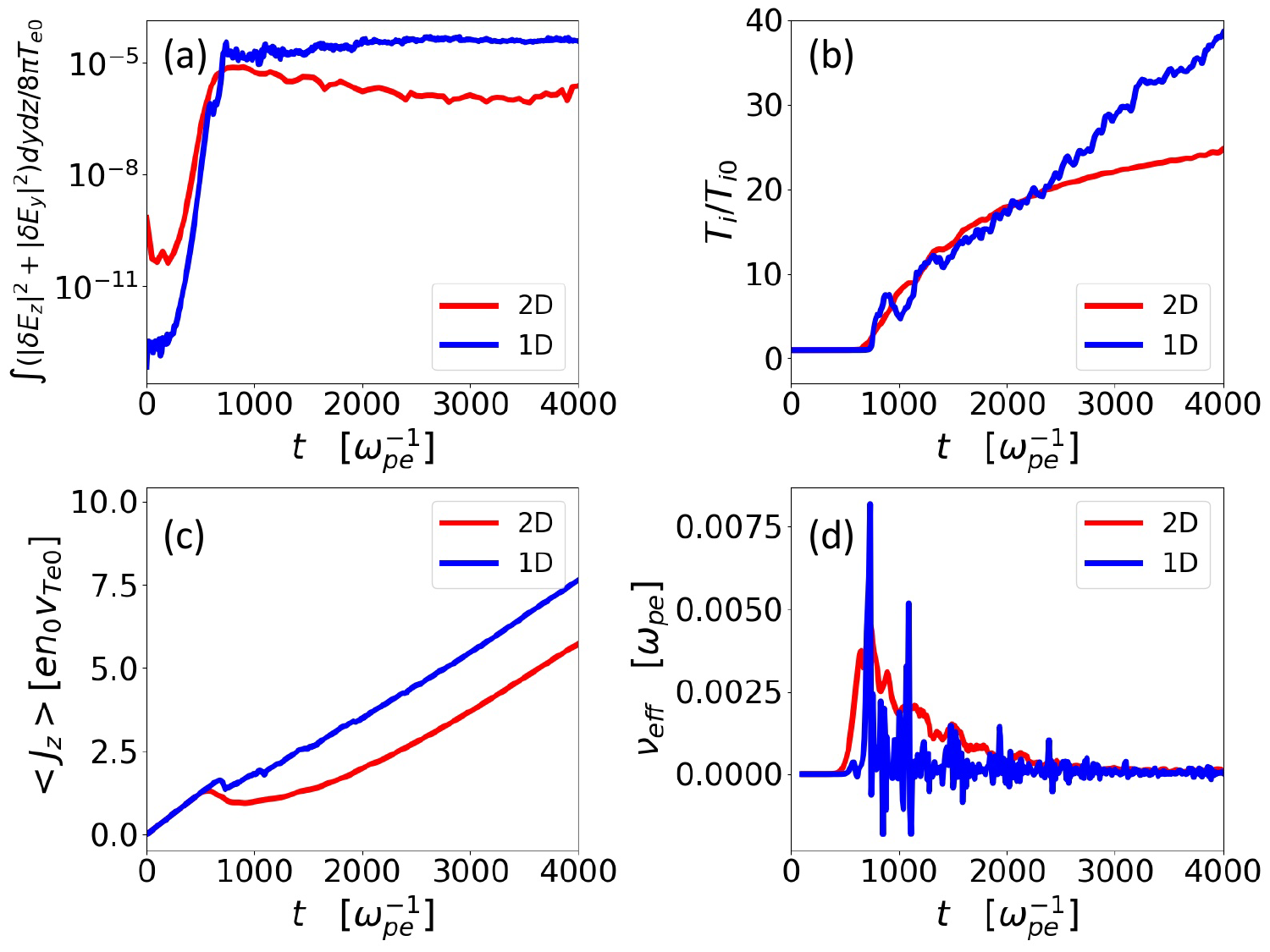}
\caption{Time traces of wave energy (a), ion temperature (b), current in the parallel direction (c), and effective collision frequency (d) from 1D and 2D simulations with the same physical parameters.}
\label{sup:fig:1D}
\end{figure}

We run a 1D1V simulation with the same physical parameters as Run {\tt Main}. 
Figure~\ref{sup:fig:1D} shows the differences between the 1D and 2D simulations. The simulation with the label of ``2D'' is Run {\tt Main}.
The evolution of the 1D system is close to the scenario described by the quasi-linear theory in that it exhibits relatively steady wave energy after saturation, as shown by the top left panel in the figure.
As a result, ions can constantly gain energy from the waves at an approximately constant rate via nonlinear trapping. We indeed see ion temperature growing linearly on the top right panel.
Moreover, as shown by the bottom left panel, although a sharp drop in the current happens during saturation, its growth rate quickly returns to the free-acceleration level. 
The corresponding anomalous resistivity, plotted in the bottom right panel, shows only a sharp peak around saturation. 
No shutdown of IAT nor transition into EAWs is observed in the 1D case,  which can be due to the different retained plateau shapes of the electron velocity distribution as explained below.

\begin{figure}
  \centering
  \includegraphics[width=0.7\textwidth]{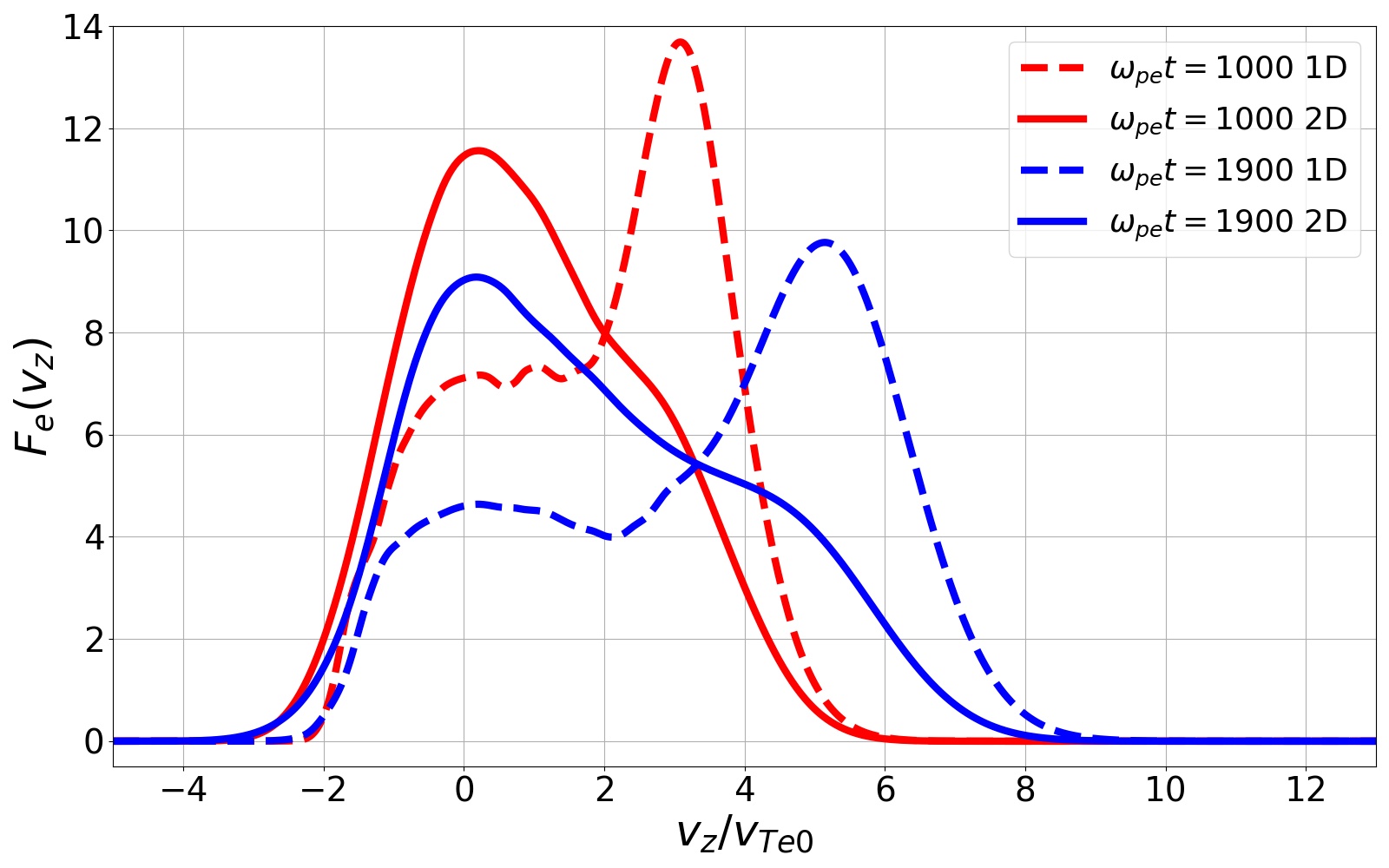}
\caption{1D electron velocity distribution function, $F_e(v_z)$ obtained from 1D (dashed curves) and 2D (solid curves) simulations with the same physical parameters. The first time moment plotted (red curves) is right after saturation, and the second time moment (blue curves) is at the end of phase III for the 2D simulation.}
\label{sup:fig:1D_dist}
\end{figure}

We plot 1D electron velocity distribution functions in Figure~\ref{sup:fig:1D_dist} right after saturation ($\omega_{pe}t=1000$) and at the end of phase III for the 2D case ($\omega_{pe}t=1900$).
In the 1D case, after saturation, the electron distribution remains flat at the phase velocities of the waves.
This plateau greatly reduces the wave emissions by electrons and keeps the wave energy approximately constant afterward (see the red and blue dashed curves in Figure~\ref{sup:fig:1D_dist}).
On the other hand, the bulk of the electrons is still located on the right-hand side of the plateau and are still freely accelerated by the external electric field.
Therefore, after the quick drop in current due to plateau formation, the current growth rate quickly returns to a value that is close to the free acceleration rate.
On the contrary, in the 2D case, plateau formation in the parallel direction cannot shut down the growth of IAWs due to the presence of oblique modes.
The saturation of these oblique modes makes the scattering much more efficient than the 1D case, and scatters most of the electrons back to the resonance region (the region with lower $v_z$).
Consequently, the 1D electron distribution exhibits in the 2D case a bulk-tail shape, where both bulk and tail can be reasonably well approximated with Maxwellian functions (see the red and blue solid curves in Figure~\ref{sup:fig:1D_dist}). As the bulk remains stationary after saturation, the current growth rate is significantly reduced.
In sum, the qualitative and quantitative differences between the electron distribution functions obtained in the 1D and 2D runs result in very different nonlinear dynamics and demonstrate that the physics of IAT is not well captured by 1D models.

\section{Electron velocity distribution fitting parameters~\label{app:2}}
We summarize the fitting parameters at different moments for Run {\tt Main} in table~\ref{sup:tab:fitted}.

 \begin{table}
\renewcommand\arraystretch{1.2}
\setlength{\tabcolsep}{12pt}
\begin{center}
\begin{tabular}{c c c c c c}
\hline
  $t\omega_{pe} $ & $\alpha$ & $v_{Te,\rm{b}}/v_{Te0}$ & $u_{d,\rm{b}}/v_{Te0}$ & $v_{Te,\rm{t}}/v_{Te0}$ & $u_{d,\rm{t}}/v_{Te0}$ \\
\hline
750  &  0.38 & 1.48 & 0.06 & 1.28 & 2.40 \\
1000  & 0.38 & 1.55 & 0.01 & 1.75 & 2.44 \\
1600  & 0.45 & 1.63 & 0.02 & 2.39 & 3.07\\
1900  & 0.56 & 1.63 & 0.05 & 2.83 & 3.23\\
\hline
\end{tabular}
\caption{ Fitting parameters of the double Maxwellian model, Eq.~\eqref{eq:double_max}, for Run {\tt Main} at different times.}
\label{sup:tab:fitted}
\end{center}
\end{table}

\begin{figure}
  \centering
  \includegraphics[width=0.6\textwidth]{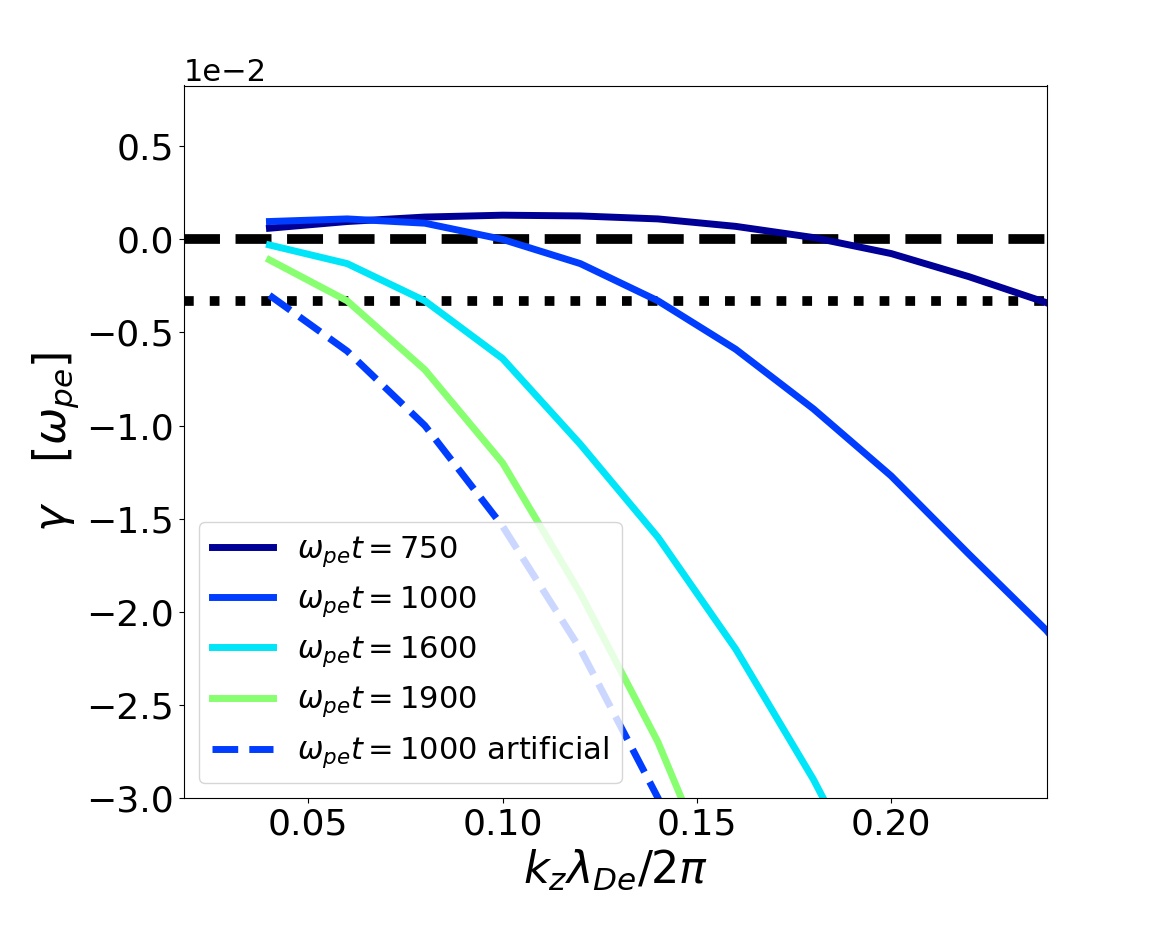}
\caption{\label{sup:fig:7.omegak_fitted} Growth rates of different wave modes obtained by solving Eq.~\eqref{eq:dielectric} at different times. The electron distribution function is represented by the double Maxwellian model, Eq.~\eqref{eq:double_max}, with the fitted parameters summarized in table~\ref{sup:tab:fitted}. 
Ion distribution is a single Maxwellian function with the ion temperature and the drift velocity at the corresponding time. The blue dashed curve is obtained by artificially increasing the ion temperature to that at $\omega_{pe}t=1900$. The black dashed marks $\gamma = 0$. The black dotted line marks the relatively strong damping rate during phase III ($\gamma = - 0.003\omega_{pe}$). }
\label{sup:1D}
\end{figure}

The growth rates of different wave modes at different times, obtained by solving the linear dispersion relation, are plotted in Figure~\ref{sup:fig:7.omegak_fitted}.
It can be seen that all the wave modes have growth rates close to zero at the moment of saturation ($\omega_{pe}t=750$).
However, as the electron-to-ion temperature decreases, wave modes begin to be more strongly damped, especially the ones with larger wavenumbers.
The blue dashed curve is obtained by using the fitted electron distribution at $\omega_{pe}t=1000$ but artificially increasing the ion temperature to its value at $\omega_{pe}t=1900$; this yields a similar curve to that obtained at $\omega_{pe}t=1900$.
Therefore, we conclude that the strong damping of IAWs is indeed mainly caused by ion heating.

\section{Electron-acoustic wave (EAW) dispersion relation~\label{app:3}}

To identify the wave modes in phase V, we calculate the theoretical dispersion relation of EAWs and compare it with the observed wavenumbers and frequencies in our simulation.
For electron waves, the Landau dispersion relation takes the form~\citep{Landau1946}
\begin{equation}
\label{sup:eq:landau}
1-\frac{1}{k^{2}} \int_{\mathrm{L}} d v \frac{\partial F_{0} / \partial v}{v-\omega / k}=0,
\end{equation}
where $\omega = \omega_{r} + i\gamma$ is the complex frequency, $k$ the wavenumber,
and $F_0$ the electron velocity distribution in the direction of wave propagation.
The EAW dispersion relation is obtained by retaining only the principal part of the velocity integral in Eq.~\eqref{sup:eq:landau}, i.e.,
\begin{equation}
\label{sup:eq:eaw}
1-\frac{1}{k^{2}} \mathrm{P}\int_{-\infty}^{+ \infty} d v \frac{\partial F_{0} / \partial v}{v-\omega / k}=0.
\end{equation}
This is because EAW theory assumes that there is a narrow plateau in the electron velocity distribution around the phase velocities of the EAWs, thereby rendering the waves immune to Landau damping~\citep{holloway1991undamped}.
The assumption can be valid because the trapped particle distribution can flatten the velocity distribution at the wave phase velocity.

\begin{figure}
  \centering
  \includegraphics[width=1.0\textwidth]{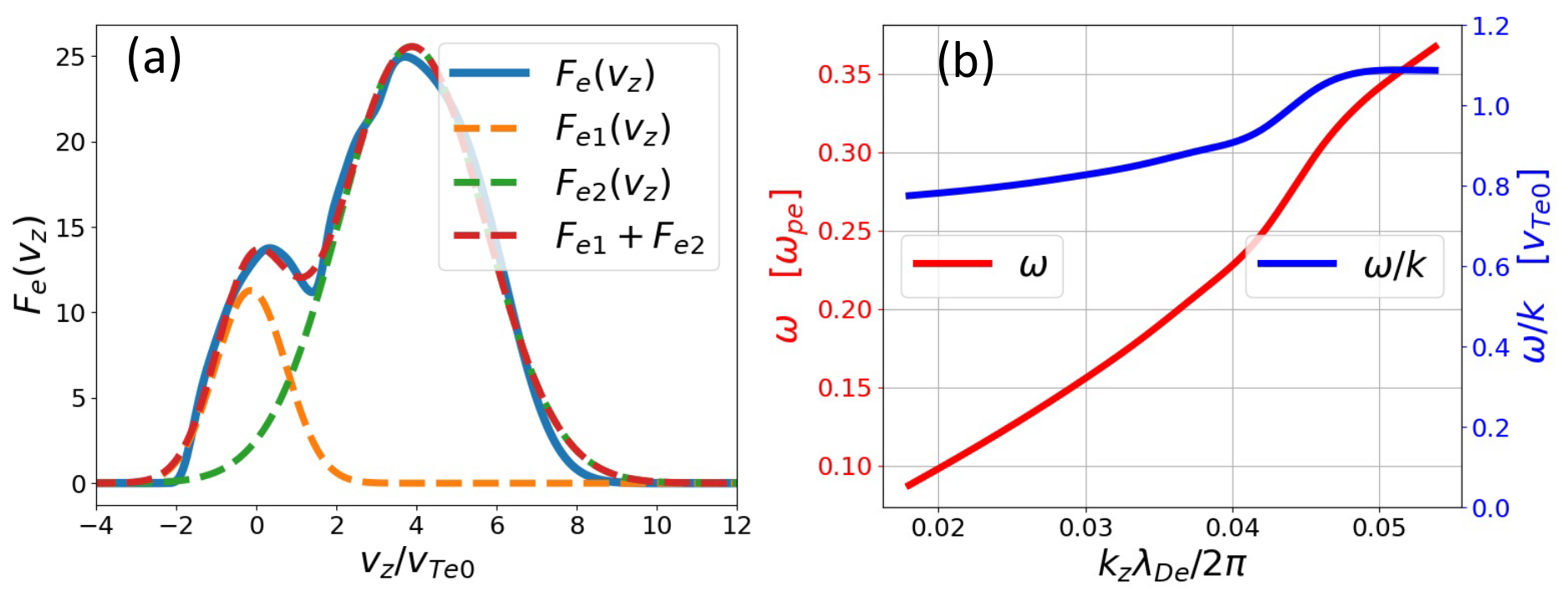}
\caption{(a) 1D electron velocity distribution and its fitting result using a double Maxwellian model. The electron distribution snapshot is taken during the burst of EAWs in phase V from Run {\tt M25E10}. (b) Dispersion relation obtained by solving Eq.~\eqref{sup:eq:eaw} with the double Maxwellian fit shown in (a).}
\label{sup:fig:eaw}
\end{figure}

To solve Eq.~\eqref{sup:eq:eaw} analytically, we consider parallel wave modes and use the double Maxwellian model to approximate the 1D electron distribution during phase V of Run {\tt M25E10}. 
The fitted result is shown in Figure~\ref{sup:fig:eaw}(a), and the dispersion relation obtained by solving Eq.~\eqref{sup:eq:eaw} is plotted in Figure~\ref{sup:fig:eaw}(b).
These theoretically predicted wavenumbers, frequencies, and phase velocities agree well with the observed properties of waves during phase V (see the $\omega-k$ diagram in Figure~\ref{fig:7.omegak}).

\section{Extremely weak field regime \label{app:4}}
The threshold external electric field of the {\it extreme weak field regime} can be estimated as follows.
The growth rate of IAI predicted by the linear theory is approximately
\begin{equation}
    \gamma (k, t) = \sqrt{\frac{\pi m_e}{8m_i}} ku (t),
\end{equation}
where $u(t) = eE_{\rm{ext}}t/m_e$ is the bulk electron drift velocity~\citep{schekochihin2022lectures}.
The wave energy density, $W$, as a function of time is then 
\begin{equation}
    W(t) =  W_0 \exp{ \left(2k \sqrt{\frac{\pi m_e}{8m_i}} \frac{eE_{\rm{ext}}}{m_e} t^2\right)},
\end{equation}
where $W_0$ is the initial wave energy density (noise).
The wave energy density at the moment of saturation, $W_{\rm{sat}}$, can be estimated by quasi-linear theory; it is 
\begin{equation}
\label{sup:eq:wsat}
    W_{\rm{sat}} \approx n_0 T_{e0} \left( \frac{E_{\rm{ext}}^2}{8\pi n_0 T_{e0}} \frac{m_i}{m_e}\right)^{1/2}.
\end{equation}
The time at which saturation occurs, i.e., $t_{\rm{sat}}$, can thus be estimated by solving
\begin{equation}
\label{sup:eq:tsat_eq}
    W(t_{\rm{sat}})  \approx  W_{\rm{sat}}.
\end{equation}

We can estimate the drift velocity of electrons in the tail at the moment of saturation
\begin{equation}
    u_{\rm{sat}} \approx c_s + \frac{e E_{\rm{ext}}}{m_e}t_{\rm{sat}}.
\end{equation}

To be in the {\it extremely weak field regime}, the difference between the ion sound speed (phase velocity of the waves) and the drift velocity of the electron tail at the moment of saturation should be smaller than the resonance width so that most of the tail can be effectively trapped by the waves, i.e., 
\begin{equation}
    \label{sup:eq:usat}
    u_{\rm{sat}} - c_s  \approx \frac{e E_{\rm{ext}}}{m_e}t_{\rm{sat}} \lesssim \frac{1}{2}  \Delta v_{\rm{trap}},
\end{equation}
where $\Delta v_{\rm{trap}}$ is the velocity width (in the parallel direction) of an electron hole in phase space (i.e. trapping width).
$\Delta v_{\rm{trap}}$ can be estimated using 
\begin{equation}
    e\phi \sim m_e \Delta v_{\rm{trap}}^2,
\end{equation}
where $\phi$ is the typical electric potential at the moment of saturation in the system.
We can estimate $\phi$ using the saturation wave energy density, i.e.
\begin{equation}
     \frac{(k \phi)^2}{8\pi} \sim \frac{E^2}{8\pi}  \sim W_{\rm sat},
\end{equation}
where we take $k$ to be the wave number of the mode with the largest linear growth rate ($k\approx 2\pi \times 0.12 \lambda_{De}^{-1}$). 
Therefore,
\begin{equation}
\label{sup:eq:phi}
\phi \sim \frac{\sqrt{8\pi n_0 T_{e0}}}{k}  \left( \frac{E_{\rm{ext}}^2}{8\pi n_0 T_{e0}} \frac{m_i}{m_e}\right)^{1/4}.
\end{equation}
The corresponding trapping width is then
\begin{equation}
\label{sup:eq:trapwidth}
   \Delta v_{\rm{trap}} \sim  \sqrt{ \frac{e \phi }{m_e} } \sim \left( \frac{8\pi n_0 T_{e0}e^2}{m_e^2 k^2}\right)^{1/4} \left(  \frac{E_{\rm{ext}}^2}{8\pi n_0 T_{e0}} \frac{m_i}{m_e} \right)^{1/8}.
\end{equation}
Substituting $t_{\rm{sat}}$ and Eq.~\eqref{sup:eq:trapwidth} into Eq.~\eqref{sup:eq:usat}, we obtain the inequality

\begin{equation}
\label{sup:eq:eext}
    \ln \left[\frac{n_0 T_{e0}}{W_0} \left( \frac{E_{\rm{ext}}^2}{8\pi n_0 T_{e0}}\frac{m_i}{m_e}\right)^{1/2}\right]  \lesssim \frac{1}{4}\left(\frac{\pi}{8}\right)^{1/2} \left(  \frac{8\pi n_0 T_{e0}}{E_{\rm{ext}}^2} \frac{m_e}{m_i} \right)^{1/4}.
\end{equation}

We now can find a critical $E_{\rm{ext}}$, i.e. $E_{\rm{ext}}^{\rm{crit}}$, that makes the approximate sign in inequality~\eqref{sup:eq:eext} hold.
Numerically, we find the $E_{\rm{ext}}^{\rm{crit}}$ under our simulation setup for the mass ratio of 25 is (assuming $W_0/n_0 T_{e0} \sim 10^{-8}$)
\begin{equation}
   \tilde{E}_{\rm{ext}}^{\rm{crit}}= \frac{E_{\rm{ext}}^{\rm{crit}}}{4\pi n_0 e \lambda_{De}} \approx 5.6 \times 10^{-4},
\end{equation}
which is consistent with our simulation results. Run {\tt M25E2} and {\tt M25E1} should indeed be in the {\it extremely weak field regime}.

\begin{figure}
  \centering
\includegraphics[width=0.9\textwidth]{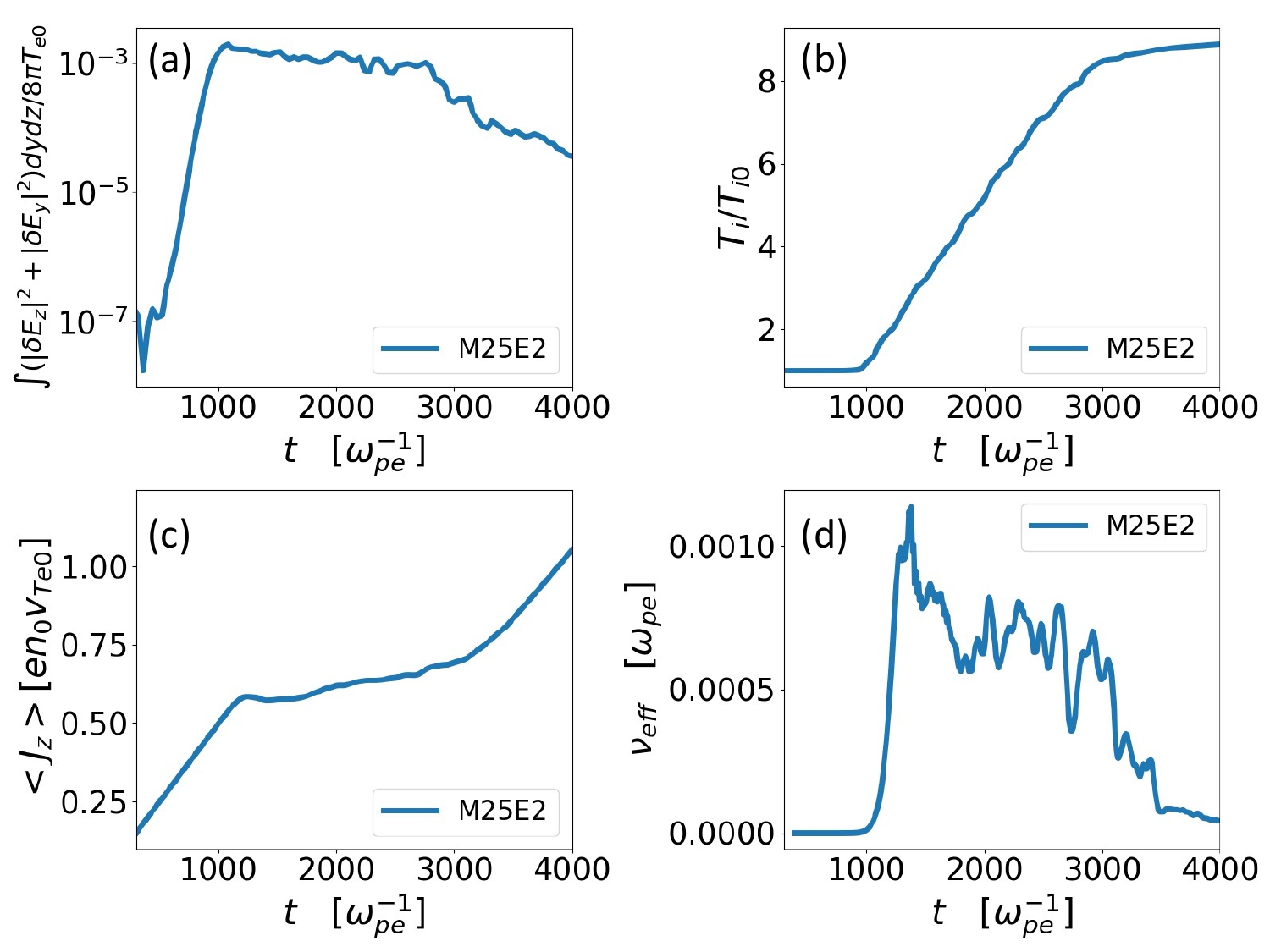}
\caption{Time traces of wave energy (a), ion temperature (b), current in the parallel direction (c), and effective collision frequency (d) from Run {\tt M25E2}. A clear shutdown moment is visible from these time traces.}
\label{sup:fig:extreme}
\end{figure}

In the {\it extremely weak field regime}, the dynamics in phase III exhibit some distinct features. 
For example, in Run {\tt M25E2}, as shown in Figure~\ref{sup:fig:extreme} (a) and (c), both the wave energy and current remain approximately constant after saturation. 
The key difference between this regime and Run {\tt Main} is that a large fraction of electrons, including those in the tails, become resonant with IAWs. As the tail portion cannot efficiently run away, the shape of electron velocity distribution barely changes during phase III.
Consequently, wave energy will not decrease until a critical temperature ratio in reached.
In this simulation, after the critical temperature ratio ($T_e/T_i \sim 10$) is reached at around $\omega_{pe}t = 3000$, the IAWs are quickly damped, leading to a rapid decrease in wave energy. 
The corresponding ion heating and anomalous resistivity, plotted in (b) and (d) of Figure~\ref{sup:fig:extreme}, are also quickly turned off.

\section{Determining the numerical resolution\label{app:5}}
\subsection{Ideal resolution\label{app:5-1}}
The numerical resolution in real space is exclusively determined by the unstable wavenumbers of IAI. 
Assuming $n$ grid points are required to resolve a single wavelength, the corresponding grid size is approximately 
\begin{equation}
\Delta x \approx \frac{2\pi}{ {k_{\text{max}}n}},
\label{sup:eq:real_space}
\end{equation}
where $k_{\rm{max}}$ is the largest unstable wavenumber of IAWs, which can be predicted from linear theory.
Conveniently, we find that $k_{\rm{max}}$ depends weakly on the drift velocity between electrons and ions according to linear theory, which implies that the ideal resolution in real space is unchanged throughout the course of the simulations. 
According to the numerical scheme we are using, $n$ is around 10.
We can see in Figure~\ref{sup:fig:linear_theory} (a) that $k_{\rm{max}}\lambda_{De} \approx  2\pi \times 0.36$ for all mass-ratio cases. 
Plugging this $k_{\rm{max}}$ into Eq.~\eqref{sup:eq:real_space}, we obtain
\begin{equation}
    \frac{\Delta x}{\lambda_{De}} \approx 0.3.
\label{sup:eq:real_ideal}
\end{equation}

With regard to velocity space, we use two methods to guide the choice of numerical resolution. 
The first method considers the phase velocities of linearly unstable IAWs.
We assume that at least three velocity grid points are required to be located in the interval of phase velocities of unstable IAWs $\left[v_{\text{ph,min}}, v_{\text{ph,max}}\right]$  i.e.,
\begin{equation}
    \Delta v_{1} \approx \frac{v_{\text{ph,max}}-v_{\text{ph,min}}}{3},
\label{sup:eq:velocity_1}
\end{equation}
This method was adopted in~\citet{petkaki2006anomalous}.

The second method estimates the width of the electron resonance region due to the IAWs when the system enters the non-linear regime. 
We may assume that a mode will enter the nonlinear regime when the bouncing frequency of an electron in the potential well caused by this wave is comparable to the wave's linear growth rate $\gamma_k$~\citep{Sagdeev1967}. 
If the maximum electric potential of the wave mode is $\phi$, the bouncing frequency is $\omega_B = \sqrt{ek^2\phi/2m_e}$. 
Letting $\omega_B \sim \gamma_{k}$, we obtain $\phi \sim m_e \gamma_{k}^2/ek^2$. 
The resonance width in electron velocity space, $\Delta v$, caused by this wave is $\Delta v \sim \sqrt{e\phi/m_e}$, and the corresponding resolution is
\begin{equation}
    \Delta v_{2} \sim \left( \frac{\gamma_{k}}{k} \right)_{\text{min}}.
\label{sup:eq:velocity_2}
\end{equation}

To resolve most of the unstable waves, estimates given by Eq.~\eqref{sup:eq:velocity_1} and Eq.~\eqref{sup:eq:velocity_2} produce similar results according to Figure~\ref{sup:fig:linear_theory} (b) and (c) (we truncate the wave number at $k\lambda_{De} = 0.3$ when estimating with Eq.~\eqref{sup:eq:velocity_2}). 
From Figure~\ref{sup:fig:linear_theory} (b) and (c), we also see that the ideal numerical resolution in velocity space depends linearly on the inverse of the square root of mass ratio.
We summarize them here:
\begin{align}
    \left(\frac{\Delta v}{v_{Te0}}\right)_{m_i/m_e=25} \approx 0.03 \label{sup:eq:velocity_ideal_1}, \\
    \left(\frac{\Delta v}{v_{Te0}}\right)_{m_i/m_e=100} \approx 0.015 \label{sup:eq:velocity_ideal_2}, \\
    \left(\frac{\Delta v}{v_{Te0}}\right)_{m_i/m_e=400} \approx 0.007 \label{sup:eq:velocity_ideal_3}.
\end{align}

\subsection{Linear benchmarks}
We perform a set of benchmarks against linear theory to check both the simulation setup and the numerical resolution. 
In these simulations, no external electric field is applied. 
Electrons have an initial drift velocity ($u_e = 0.5 v_{Te0}$, which is a typical drift velocity during phase I of most of our simulations in the paper) such that the initial condition is ion-acoustic unstable. 
An ion-to-electron mass ratio of 25 and an electron-to-ion initial temperature of 50 are used. We only perturb one wave mode in each simulation. All the simulations are run to the end of the linear stage.

The numerical resolution we use in our simulations differs from Eq.~\eqref{sup:eq:real_ideal} and Eq.~\eqref{sup:eq:velocity_ideal_1}.
This is because adopting Eq.~\eqref{sup:eq:real_ideal} demands unaffordable computational power.
However, due to the fact that modes with very small or large wavenumbers are weak in amplitude and thus unimportant to the nonlinear evolution of the system, we only need to resolve modes within a certain range of wavenumbers that have relatively strong growth rates. 
Several different resolution combinations are proposed in table~\ref{table:combination} based on the estimates given by \eqref{sup:eq:real_ideal} and \eqref{sup:eq:velocity_ideal_1}. 
We refer to grid sizes in real space as  ($\Delta z, \Delta y$) and grid sizes in electron and ion velocity spaces as ($\Delta v_{e,z}, \Delta v_{e,y}$, $\Delta v_{i,z}$, $\Delta v_{i,y}$).
The combination \#3-1 has the same resolution in the $z$ (parallel) direction with \#3, and it is only used for the nonlinear test described in section~\ref{app:5-3}.
The ion velocity domain is small and thus demands much fewer computational resources. 
We set $\Delta v_{i,z}  = \Delta v_{i,y} = v_{Ti0}/3$ for all simulations.

\begin{table}
\renewcommand\arraystretch{1.2}
\setlength{\tabcolsep}{12pt}
\begin{center}
\begin{tabular}{c c c}
\hline
  No. & $\{\Delta z,\Delta y\}/\lambda_{De}$ &$ \{\Delta v_{e,z}, \Delta v_{e,y}\}/  v_{Te0}$ \\
\hline
1  & $\{1.0 , 1.0\}$  & $\{0.042, 0.042\}$\\
2  & $\{0.5 , 0.5\}$  & $\{0.083, 0.083\}$\\
3  &  $\{0.5 , 0.5\}$  & $\{0.042, 0.042\}$ \\
3-1  & $\{0.5 , 0.5\}$  & $\{0.042, 0.083\}$ \\
4  &  $\{0.5 , 0.5\}$  & $\{0.021, 0.042\}$\\
\hline
\end{tabular}
\caption{Resolution combinations for linear benchmarks and nonlinear tests.}
\label{table:combination}
\end{center}
\end{table}

\begin{figure}
  \centering
\includegraphics[width=0.8\textwidth]{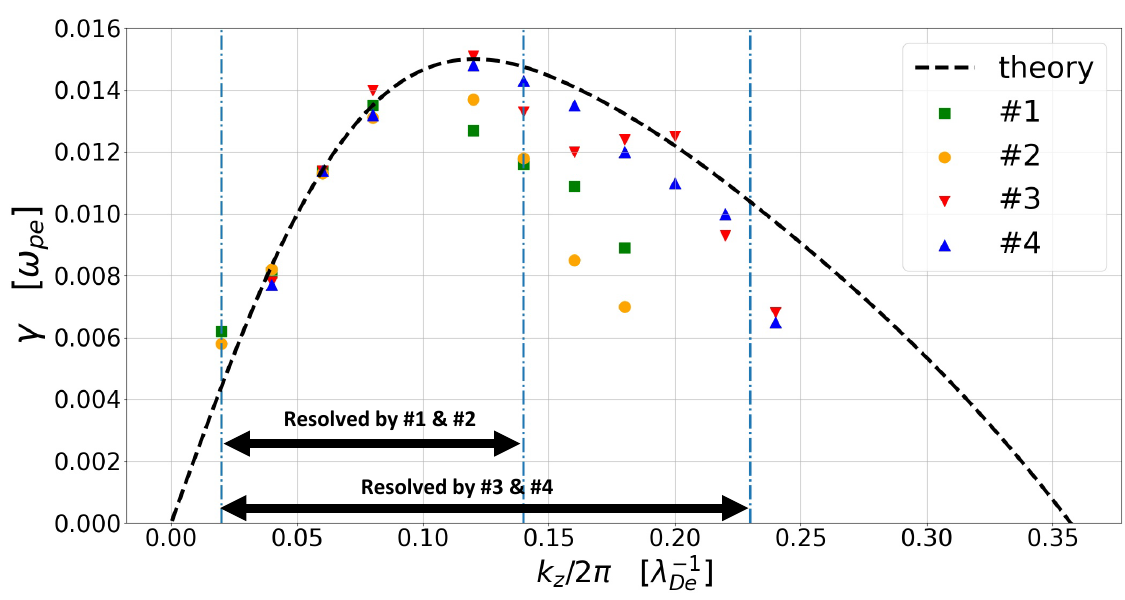}
\caption{Linear benchmark results. The black dashed curve is the solution to the ion-acoustic analytical dispersion relation Eq.~\eqref{sup:eq:dieletric} for $m_i/m_e=25$ and $T_i/T_e=50$ with electron drift velocity of $u_e=0.5v_{Te0}$. Symbols are obtained from numerical simulations performed with the numerical resolution combinations indicated in table~\ref{table:combination}.}
\label{sup:mass25_linear}
\end{figure}

The benchmark results are concluded in Figure~\ref{sup:mass25_linear}. 
Combination \#1 can correctly resolve the growth rates with wavenumbers smaller than about $2\pi\times 0.14 \lambda_{De}^{-1}$, while combination \#3, and \#4 can resolve higher wavenumber modes up to $2\pi\times 0.22 \lambda_{De}^{-1}$.
Different from combination \#1, combination \#2 cannot resolve wave modes with higher wavenumbers because of insufficient numerical resolution in electron velocity space.
The fact that combinations \#3 and \#4 can resolve a similar range of wave modes implies that  $\Delta v_{e,z} = 0.042 v_{Te0} $ is enough to resolve most of the linearly unstable wave modes.

\subsection{Nonlinear tests\label{app:5-3}}

\begin{figure}
  \centering
\includegraphics[width=0.9\textwidth]{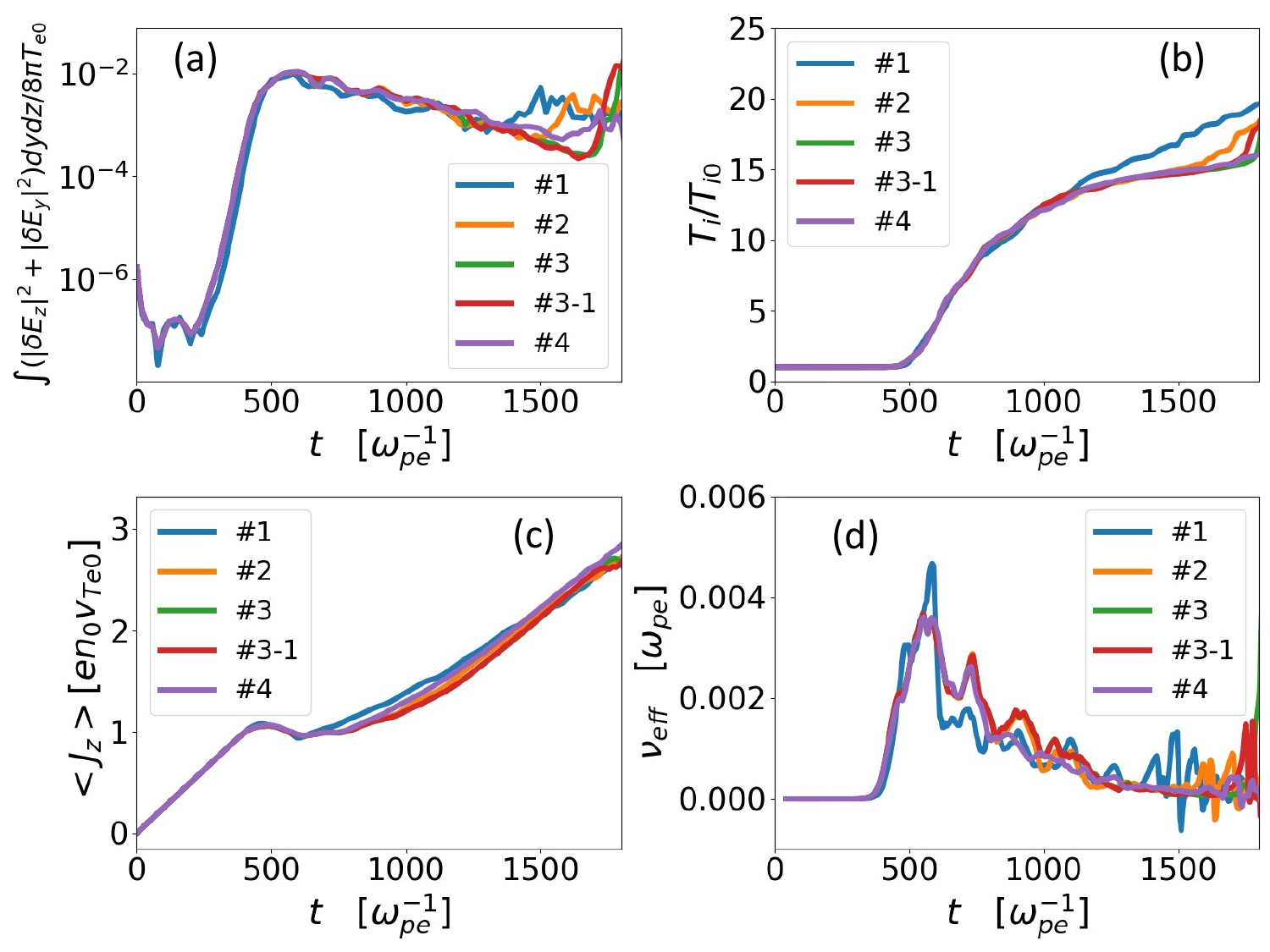}
\caption{
Time traces of wave energy (a), ion temperature (b), current in the parallel direction (c), and effective collision frequency (d) from nonlinear evolution tests with different numerical resolution combinations (indicated in table~\ref{table:combination}).}
\label{sup:fig:mass25_nonlinear}
\end{figure}

To understand how these numerical resolutions influence the nonlinear evolution of our system, we perform turbulence simulations with different numerical resolution combinations in table~\ref{table:combination}. 

The setup for these turbulent tests is the same as the results presented in the main paper. Both electrons and ions have zero drift velocity at the start, and an external electric field is applied to drive IAI.
The external electric field is chosen to be $\tilde{E}_{\rm{ext}} =2.5\times 10^{-3}$, which corresponds to {\tt E10} in run names. All simulations are long enough to identify the bursts of EAWs. 

The time traces for perturbed electric field energy, ion temperature, current in the parallel direction, and anomalous resistivity are plotted in Figure~\ref{sup:fig:mass25_nonlinear}. 
It can be seen that all of these resolution combinations exhibit similar linear and nonlinear evolution behavior.

The simulation that deviates most from the others is combination \#1: 
it exhibits higher ion heating than the others and a somewhat shorter phase III and phase IV. 
On the other hand, combination \#2 behaves much better than resolution combination \#1, even though it is only able to resolve a similar wavenumber range as combination \#1 in the linear benchmarks. 
As expected, not being able to resolve the linear growth rates of some modes correctly does not imply that the nonlinear behavior will also be substantially influenced because wave amplitude is small during the linear phase and thus requires finer grids to resolve it.
This observation implies that lower resolution in velocity space is less impactful in the nonlinear stage.
In addition, the highly overlapped time traces between combination \#3 and \#3-1 imply that lower resolution in the $y$ direction of electron velocity space is harmless.
Based on these observations, we finally conclude that resolution combination \#3-1 can capture nonlinear physics accurately enough for the mass ratio of 25 cases.

To summarize, as the real space numerical resolution does not depend on mass ratio, the grid size for all the simulations can be chosen as
\begin{equation}
        \{\Delta z, \Delta y\} = \{0.5 , 0.5\}\lambda_{De}.
\end{equation}
The required velocity space numerical resolution linearly depends on the inverse square root of mass ratio (See Figure~\ref{sup:fig:linear_theory} and Eq.~(\ref{sup:eq:velocity_ideal_1}-\ref{sup:eq:velocity_ideal_3})), so the resolution of electron velocity space for the other cases can be calculated based on combination \#3-1.
For example, the numerical resolution of electron velocity space of Run {\tt Main} should be 
\begin{equation}
\label{sup:eq:v_elc_res}
        \{\Delta v_{e,z}, \Delta v_{e,y}\}  = \{0.021, 0.042 \} v_{Te0}.
\end{equation}

For Run {\tt M200E10}, we are forced by computational resource constraints to keep this numerical resolution (Eq.~\eqref{sup:eq:v_elc_res}).
While insufficient to adequately capture the linearly unstable spectrum, these nonlinear tests demonstrate that phases III-IV of the evolution should still be reasonably resolved.

\bibliographystyle{jpp}

\bibliography{jpp-instructions}

\end{document}